\documentclass[a4paper,fleqn,usenatbib]{mnras}

\usepackage{newtxtext,newtxmath}
\usepackage[T1]{fontenc}
\usepackage{ae,aecompl}
\usepackage{graphicx}
\usepackage{amsmath}
\usepackage{amssymb}
\usepackage{tablefootnote}

\title[Intra-day variability of S5 0716+714]
{Intra-day optical multi-band quasi-simultaneous observation of BL Lacertae
object S5 0716+714 from 2013 to 2016}
\author[X. Zhang et al.]{
Xiaoyuan Zhang,$^{1}$   
Jianghua Wu$^{1}$\thanks{E-mail: jhwu@bnu.edu.cn} and 
Nankun Meng$^{1}$
\\
$^{1}$Department of Astronomy, Beijing Normal University, Beijing 100875,
China
}

\date{Accepted. Received; in original form}
\pubyear{2018}

\begin{document}
\label{firstpage}
\pagerange{\pageref{firstpage}--\pageref{lastpage}}
\maketitle

\begin{abstract}
We perform quasi-simultaneous optical multi-band monitoring of BL Lac object S5 0716+714 on seven nights from 2013 to 2016. Intra-day variability (IDV) is found on all seven nights. The source was faintest on JD 2456322 with 14.15 mags and brightest on JD 2457437 with 12.51 mags in the $R$ band. The maximum intra-day variation we observed is 0.15 mags in the $B$ band on JD 2456322. Both bluer-when-brighter and achromatic spectral behaviours were observed on the intra-day timescale. On the longer-term scale, the object exhibited a mild bluer-when-brighter behaviour between the $B$ and $R$ bands. We estimate the inter-band lags using two independent methods. The variation in the $B$ band was observed to lag that in the $I$ band by about 15 minutes on JD 2457315. We compare this lag with one reported previously and discussed the origin of these lags.
\end{abstract}
\begin{keywords}
galaxies: active -- BL Lacertae objects: individual: S5 0716+714 -- galaxies:
photometry
\end{keywords}

\section{Introduction}
Blazars are a subset of Active Galactic Nuclei (AGNs). They are those AGNs with their relativistic jets pointing at a small angle to our light of sight \citep{1995PASP..107..803U}. Blazars can be classified into BL Lacertae objects and flat-spectrum radio quasars (FSRQs) according to the strength of their emission lines. BL Lac Objects have absent or weak emission lines (EW$\le5$\AA), while FSRQs show strong emission lines in their spectra. In general, there are two humps in the spectral energy distribution (SED) of blazars. The first hump extending from radio to UV or soft X-ray, is likely dominated by synchrotron radiation from the relativistic jet, and the second, covering from UV or soft X-ray to $\gamma$-ray, dominated by inverse Compton emission. According to the frequency of the synchrotron peak, blazars are classified as low synchrotron peaked blazar (LSP, $\nu_{\mathrm{syn\_peak}}\le10^{14}$ Hz), intermediate synchrotron peaked blazar (ISP, $10^{14}$ Hz$<\nu_{\mathrm{syn\_peak}}<10^{15}$ Hz) and high synchrotron peaked blazar (HSP, $\nu_{\mathrm{syn\_peak}}\ge10^{15}$ Hz). The most striking characters of blazars is its dramatic variability from radio to $\gamma$-ray (e.g. \citealt{2003ApJ...596..847B,2008A&A...491..755R,2009A&A...501..455V}). The variability timescales vary from minutes to years. 
The rapid variability with timescale less than one day is called intraday variability. The short variation timescales limit the emission regions within extremely small sizes. Studying the optical intraday variability can help constraining the physical processes at the base of the blazar jets, e.g. particle acceleration and cooling mechanism, magnetic field geometry, plasma instability in the jet etc. 
Previous studies of blazar optical IDV have made great progress since the first optical IDV discovered by \citet{1989Natur.337..627M} in BL Lacertae. Systematic optical IDV search was performed by \citet{1996A&A...305...42H}, where IDV was detected in 28 out of 34 1 Jy catalog  BL Lac objects. \citet{2005A&A...440..855G} reported that the probability of IDV detection is 80 to 85\% if the blazar is continuously observed over six hours. Most of the IDV detected blazars are LSPs and ISPs (e.g. \citealt{2008AJ....135.1384G,2012MNRAS.425.3002G}), whereas HSPs have few evidence of IDV (e.g. \citealt{2012AJ....143...23G,2012MNRAS.420.3147G}). Recently, \citet{2018Galax...6....1G} has reviewed multi-wavelength IDVs of blazars. 

S5 0716+714 (RA = 07:21:53.45, Dec = 71:20:36.36, J2000), one of the brightest BL Lac objects in the northern sky, is classified as an ISP according to its synchrotron peak frequency $10^{14.6}$ Hz \citep{2010ApJ...716...30A}. The redshift is $z=0.31\pm0.08$, estimated by \citet{2008A&A...487L..29N} by using the host galaxy as a ``standard candle'', and later limited by \citet{2013ApJ...764...57D} as $z<0.322$. It is also one of the best-studied blazars with high variability from radio to $\gamma$-ray (e.g. \citealt{2008A&A...481L..79V,2013A&A...552A..11R,2014ApJ...783...83L}). In the optical regime, it exhibits fast variability with the duty cycle approximate to 1 \citep{1996AJ....111.2187W}. A number of campaigns were performed to study IDV properties of this source (e.g. \citealt{2000A&A...363..108V,2003A&A...402..151R,2005AJ....130.1466N,2005AJ....129.1818W,2007AJ....133.1599W,2012AJ....143..108W,2011AJ....141...49C,2013ApJS..204...22D,2014MNRAS.443.2940H,2016MNRAS.455..680A,2017MNRAS.469.2457L,2017AJ....154...42H}). 
\citet{2006A&A...451..435M} studied monitoring data on 102 nights from 1996 to 2003. The distribution of variability timescales followed an exponential law and the shortest timescale is about 2 hours. Seventy-two hours WEBT continuous observations show that the power spectrum density of the light curve is well fitted by $1/f^{2}$ power law, indicating the stochastic nature of the IDV \citep{2013A&A...558A..92B}. Meanwhile, quasi-periodic oscillations are occasionally reported (e.g. \citealt{2005AJ....129.1818W,2010ApJ...719L.153R,2016MNRAS.456.3168M,2016ApJ...831...92B,2018AJ....155...31H}).

During flares, spectral hystereses or time lags between two light curves at different wavelengths are sometimes observed. Most of these events are soft lags, i.e. variations at short wavelength lead that at long wavelength (e.g. \citealt{1996ApJ...470L..89T,2000ApJ...528..243K}). On the other hand, hard lags are were also observed in a few of cases (e.g. \citealt{2000ApJ...541..166F}). Different spectral hysteresis patterns as well as the position of the observation frequency relative to the synchrotron peak frequency are essential to constrain different jet models i.e. homogeneous single-zone leptonic models \citep{1998ApJ...501L.157D,1999MNRAS.306..551C} and the internal-shock model \citep{2001MNRAS.325.1559S,2010ApJ...711..445B}.
For S5 0716+714, time lags among different electromagnetic wave regimes are frequently detected. \citet{2003A&A...402..151R} reported that the radio flux variations at lower-frequencies lagged the higher-frequency ones with time delays from a few days to weeks; \citet{2013A&A...552A..11R}  stated the optical/GeV flux variations lead the radio variability by $\sim65$ days. In the optical regime, inter-band lags were reported by several authors. For example, \citet{2000PASJ...52.1075Q} reported a 6-minute lag between variations in the $V$ and $I$ bands; \citet{2000A&A...363..108V} found a 10-minute between the $B$ and $I$ bands; a plausible 11-minute lag between the $B$ and $I$ bands was observed by \citet{2009ApJS..185..511P}; recently, \citet{2012AJ....143..108W} reported a 30-minutes lag between the $B'$ and $R'$ bands; later \citet{2016MNRAS.456.3168M} observed a possible 1.5-minutes lag between the $B$ and $I$ bands. Since some of the time lags are as short as only a few minutes, high temporal resolutions are needed to increase the probability of lag detection. Therefore, we performed multi-band quasi-simultaneous observations with high temporal resolutions. In this paper, we report our observation and analysis results.

This paper is organised as follows: 
In Section \ref{obs}, we report details of observation and data reduction. In
Section \ref{results}, we show results of various analyses of our data
including IDV test, colour behaviour and
cross-correlation analysis.
Discussion and conclusion are given in Section \ref{discussion} and
\ref{conclusion}.

\begin{table*}
\centering
\caption{Parametres of telescopes and terminal instruments.}
\label{telescope}
\begin{tabular}{lccccc}
\hline
Telescope   & 2.16 m             & \multicolumn{2}{c}{85 cm}    & 80 cm      &60 cm       \\
&&Old CCD&New CCD (after 2014)&& \\
\hline
Optical Design & Ritchey - Chretien & \multicolumn{2}{c}{Prime Focus}     & Cassegrain &Prime Focus \\
Focus Ratio & f/9                & \multicolumn{2}{c}{f/3.3}   & f/10       &f/4.23      \\
CCD Model & E2V 55-30          & PI 1024 EBFT-1 & Andor     & PI 1300B   &E2V 47-10   \\
CCD Size    & 1242 $\times$ 1152  & 1024 $\times$ 1024&2048 $\times$ 2048&1340$\times$ 1300&512 $\times$ 512   \\
Pixel Scale (arcsec pixel$^{-1}$)& 0.457              &0.96 &0.96 & 0.52&1.95        \\
FOV (arcmin$^2$)        & 9.46 $\times$ 8.77 & 16.4 $\times$ 16.4 & 32.8$\times$ 32.8 & 11.5 $\times$ 11.2&16.6 $\times$ 16.6  \\
\hline
\end{tabular}
\end{table*}

\section{Observation and data reduction}\label{obs}
\subsection{Telescopes and observation strategy}
Usually, the temporal resolution of quasi-simultaneous observations by one telescope equipped with multiple filters is limited by the number of filters, because exposures with different filters are taken in a cyclic pattern. \citet{2007AJ....133.1599W} utilised an objective prism and a multi-peak interference filter to achieve exactly simultaneous observations at three passbands. However, this method may introduce some extra uncertainties when adopting an elongated aperture in photometry. Also, this method doesn't work well with a crowded stellar field. As a result, we adopted a compromising method, using multiple telescopes to monitor the object with different filters independently. It allows us to obtain high temporal resolution light curves in all bands, see \citet{2016MNRAS.456.3168M} for an example.

During our observations, four telescopes in Xinglong Observatory, National Astronomical Observatories, Chinese Academy of Science (NAOC) are used. Parameters of these telescopes are listed in Table \ref{telescope}. Computer clocks of these telescopes are synchronised by the GPS clock. Observations were performed on seven nights from 2013 to 2016. On January 29th, 2013 (JD 2456322), the 2.16 m, 85 cm and 80 cm telescopes were selected for observation. From October 16th to 19th, 2015 (JD 2457312 to JD 2457315), the 60 cm, 80 cm, 85 cm telescopes were selected. From February 17th to 18th, 2016 (JD 2457436 to JD 2457437), only the 85 cm telescope was used. Details of filters, observation durations, temporal resolutions are listed in columns 3 - 5 of Table \ref{observation}.

\begin{table*}
\centering
\caption{Details of observations.}
\label{observation}
\begin{tabular}{ccccccc}
\hline
Date & Julian Date &   Filter  &   Duration (h)    &   Temporal Resolution (s) &   Telescope  & Good Data Ratio\\
\hline
Jan. 29th 2013 & 2456322  & $B$ & 5.67 & 67.8 & 2.16 m & 0.78\\
& & $R$ & 5.50 & 119.0 & 80 cm & 0.81\\
& & $I$ & 5.33 & 23.0 & 85 cm & 0.72\\
Oct. 16th 2015 & 2457312 & $R$ & 3.51 & 50.0 & 80 cm & 0.98\\
& & $I$ & 3.39 & 45.6 & 60 cm & 1.00\\
Oct. 17th 2015 & 2457313 & $B$ & 3.18 & 200.2& 80 cm & 0.71\\
& & $V$ & 3.12 & 10.0 & 85 cm & 0.86\\
& & $R$ & 3.17 & 199.3& 80 cm & 0.60\\
Oct. 18th 2015 & 2457314 & $B$ & 2.64 & 35.0 & 85 cm & 1.00\\
& & $R$ & 2.79 & 50.0 & 80 cm & 0.99\\
& & $I$ & 2.72 & 45.6 & 60 cm & 0.91\\
Oct. 19th 2015 & 2457315 & $B$ & 3.12 & 25.2 & 85 cm & 0.99\\
& & $R$ & 3.23 & 34.0 & 80 cm &0.79\\
& & $I$ & 3.40 & 25.4 & 60 cm & 1.00\\
Feb. 17th 2016 & 2457436 & $B$ & 9.17 & 103.1& 85 cm & 1.00\\
& & $V$ & 9.15 & 103.1& 85 cm & 1.00\\
& & $R$ & 9.21 & 103.1& 85 cm & 0.99\\
Feb. 18th 2016 & 2457437 & $B$ & 4.95 & 91.1 & 85 cm & 0.74\\
& & $V$ & 4.92 & 91.1 & 85 cm & 0.82\\
& & $R$ & 4.95 & 91.1 & 85 cm & 0.80\\
\hline

\end{tabular}
\end{table*}

\subsection{Data reduction}\label{reduction}
\begin{figure}
\includegraphics[width=\columnwidth]{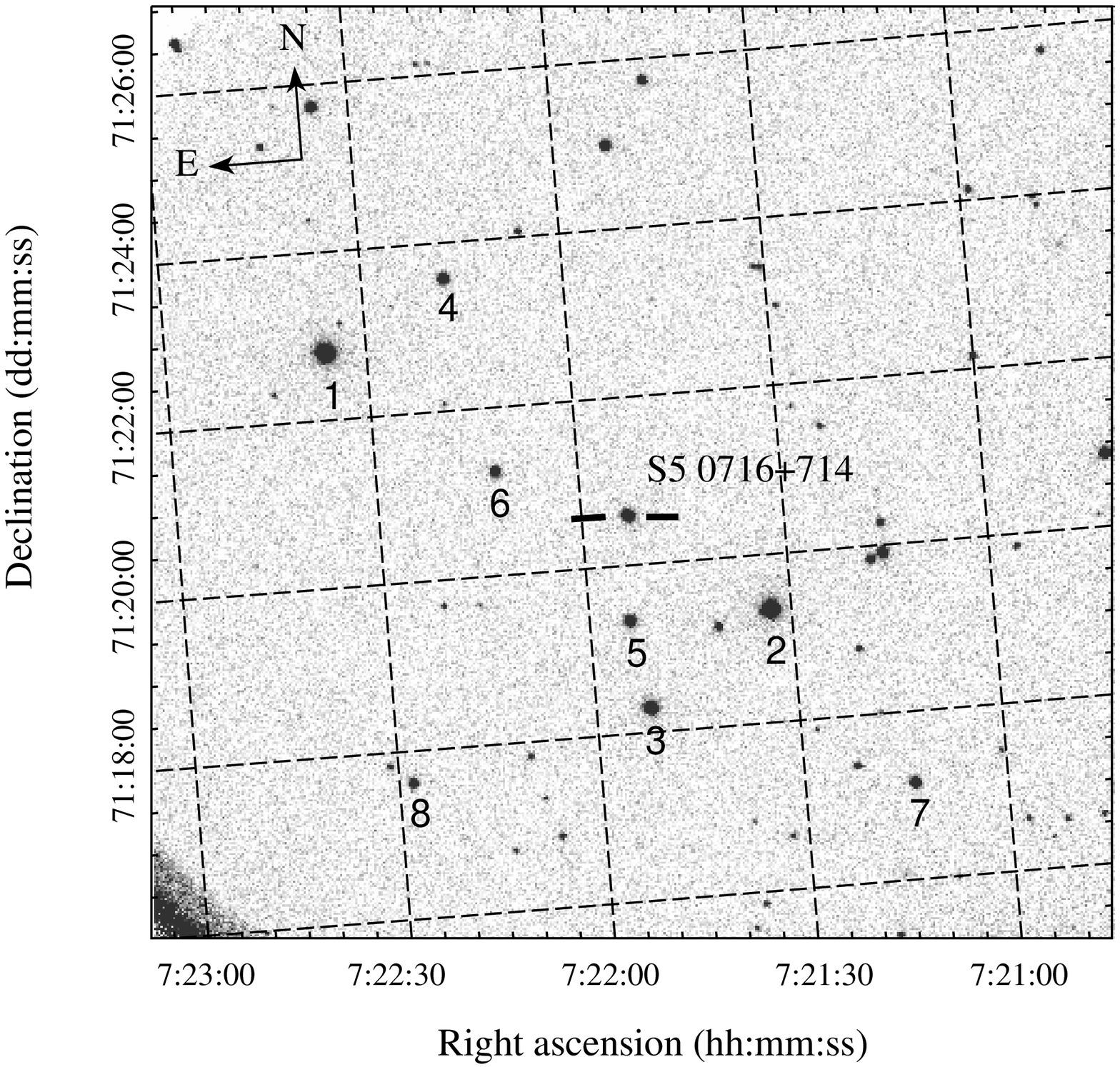}
\caption{Finding chart of S5 0716+714 in the $R$ band.}
\label{finding}
\end{figure}

For original data obtained in each session, we follow the standard process including bias-subtraction and flat-fielding with IRAF\footnote{IRAF is distributed by the National Optical Astronomy Observatories, which are operated by the Association of Universities for Research in Astronomy, Inc., under cooperative agreement with the National Science Foundation.}. To find the best aperture radius, first, we set it as 1.5, 2.0, 2.5, 3.0, 3.5 and 4.0 times of the full width at half-maximum (FWHM) of the stellar images. The inner and outer radii of the sky annuli are 7 and 9 times of the FWHM. Then the instrument magnitudes of S5 0716+714 and stars 1 to 8 in the finding chart (see Figure \ref{finding}) are extracted from the frames. In order to minimise the intrinsic error of differential magnitudes, the comparison stars should be somewhat brighter than the object \citep{1988AJ.....95..247H}. As a result, two bright unsaturated stars 2 and 3 are selected as comparison stars and star 5 as the check star. We adopt the aperture with the smallest standard deviation of differential magnitude (the difference between instrumental magnitudes of two reference stars). The magnitude of S5 0716+714 are calibrated relative to those of stars 2 and 3.  The differential magnitude of check star (the difference between instrumental magnitudes of check star and the average value of two comparison stars) is also derived to exhibit the accuracy of photometry. Standard magnitudes of all comparison stars are given by \citet{1998A&AS..130..305V} in the $B$, $V$, $R$ bands and by \citet{1997A&A...327...61G}
in the $I$ band.

\begin{figure}
\centering
\begin{tabular}{ll}
\includegraphics[width=0.45\columnwidth]{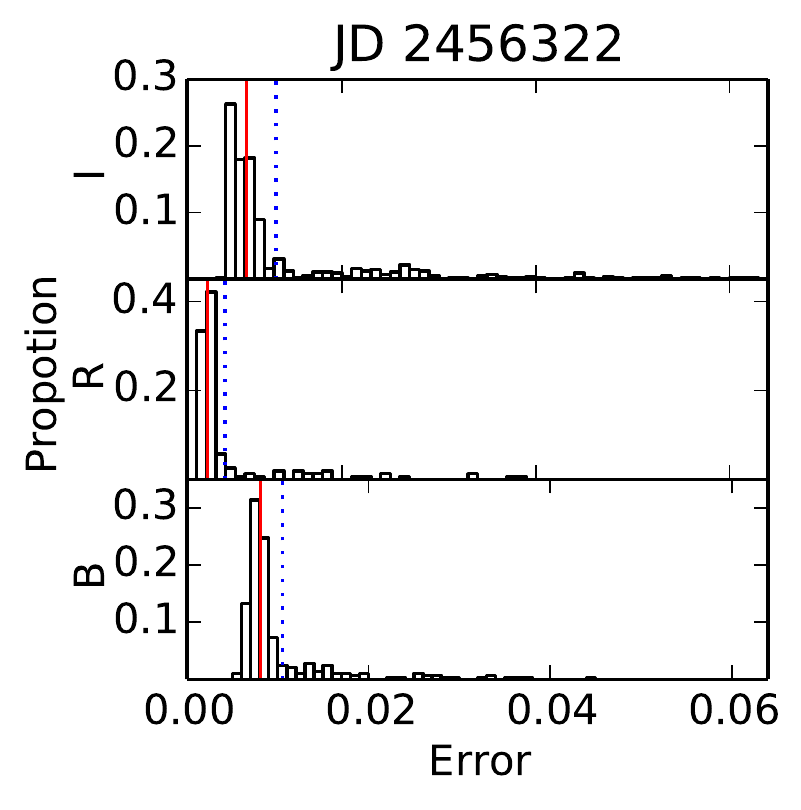}&
\includegraphics[width=0.45\columnwidth]{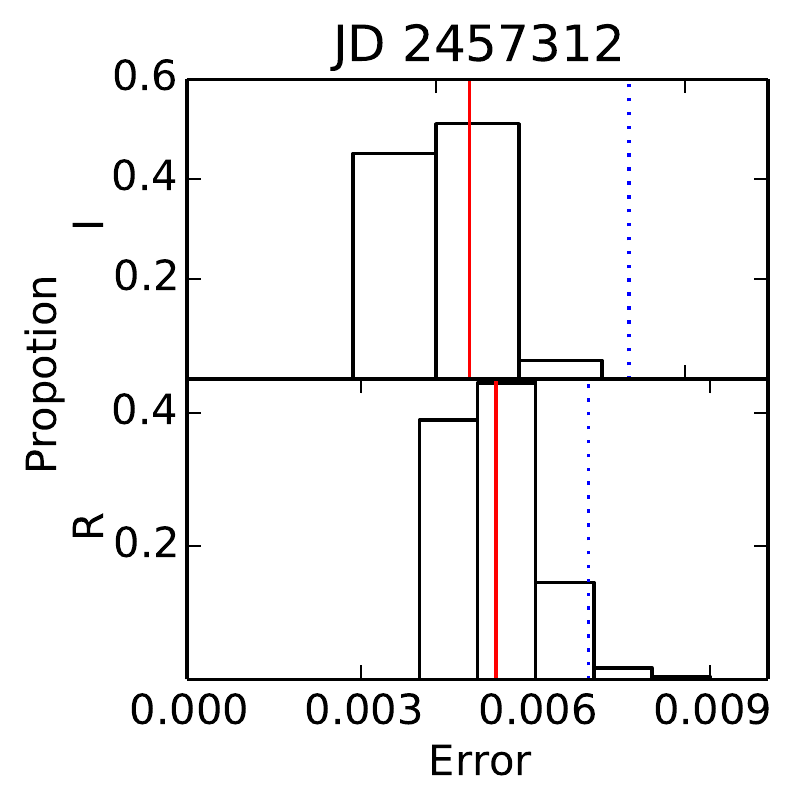}\\
\includegraphics[width=0.45\columnwidth]{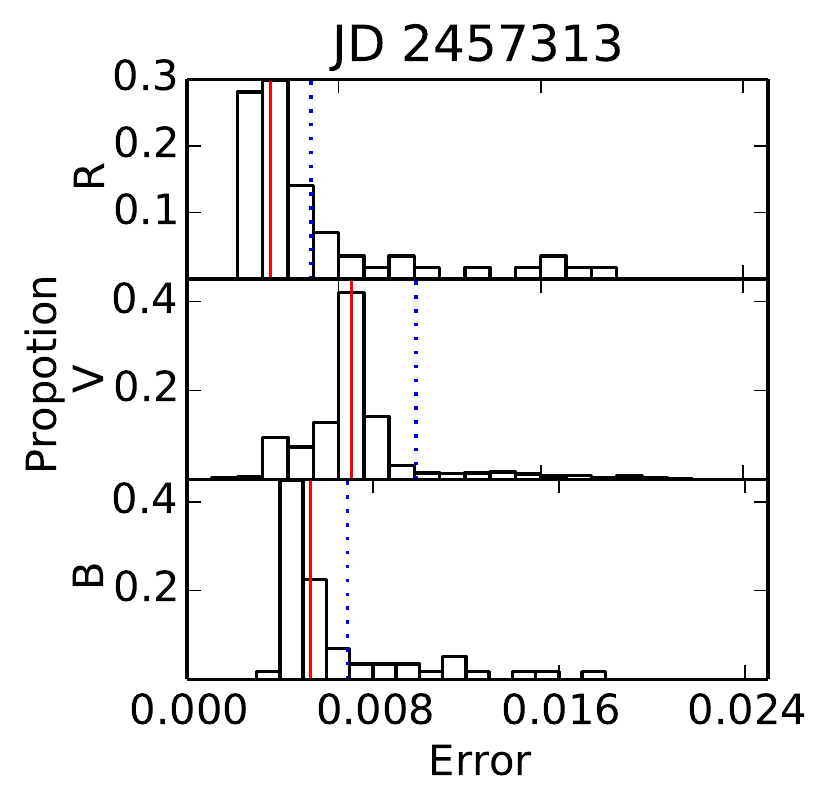}&
\includegraphics[width=0.45\columnwidth]{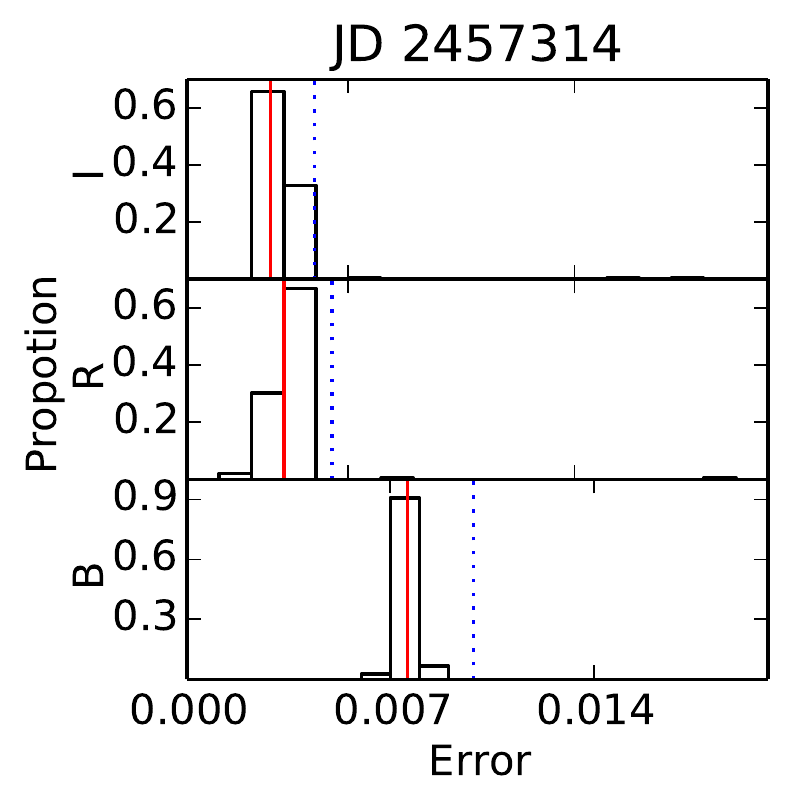}\\
\includegraphics[width=0.45\columnwidth]{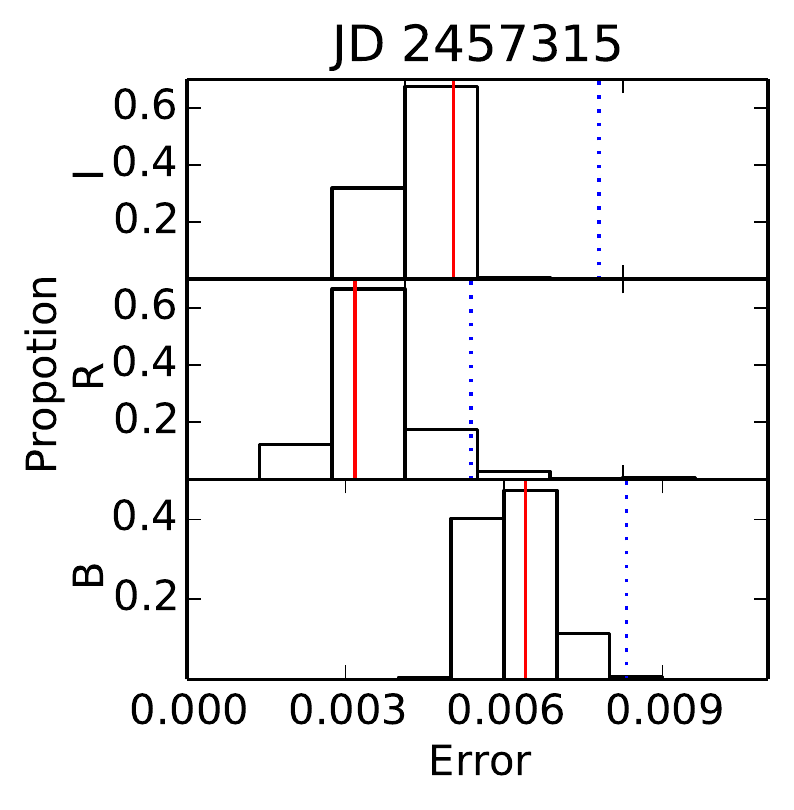}&
\includegraphics[width=0.45\columnwidth]{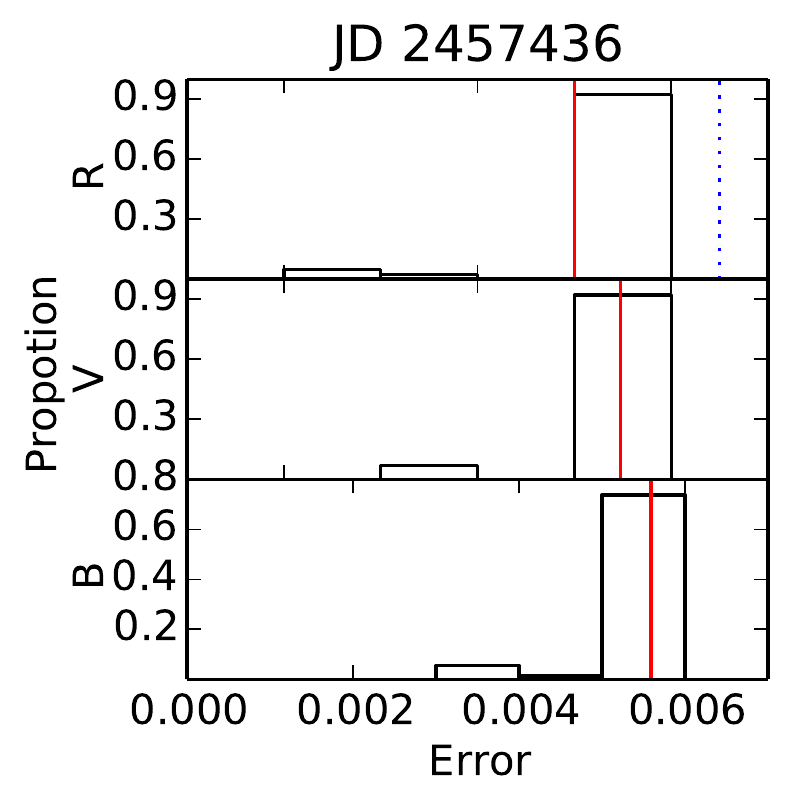}\\
\includegraphics[width=0.45\columnwidth]{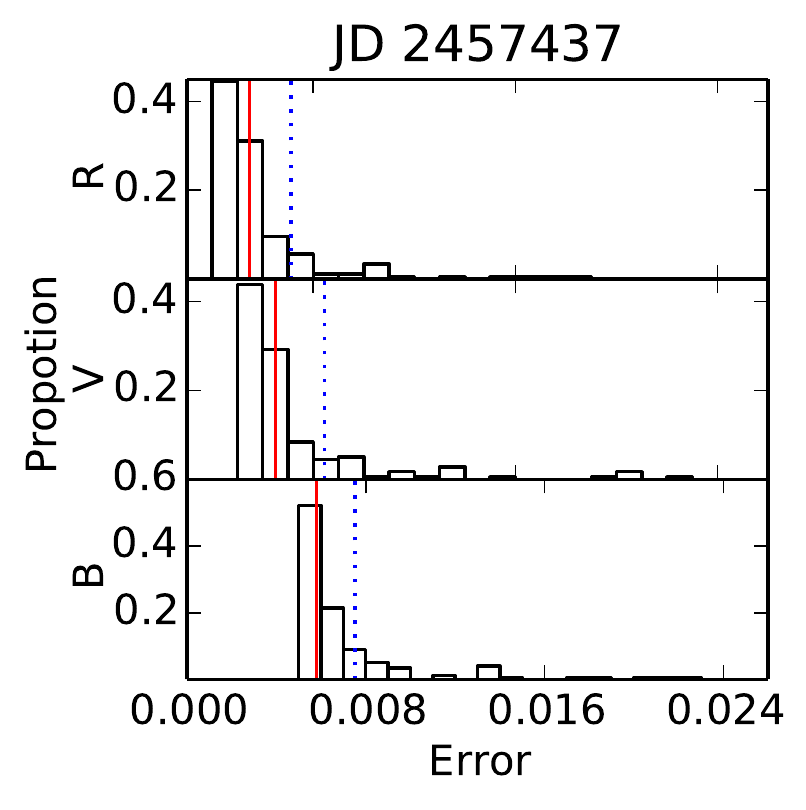}&
\end{tabular}
\caption{Photometry error distributions of each band on each night. Red solid
lines represent the median value of errors and blue dotted lines represent 1.3
times of the median value.}
\label{err_dist}
\end{figure}

We calculate the distributions of the raw photometric errors on each night. The results are plotted in Figure \ref{err_dist}. Here, the red solid and blue dotted lines represent 1 and 1.3 times of the median values of the raw photometric errors of each light curve, respectively. Errors in tails on the right side can be caused by sudden changes of weather conditions and brightening of the skylight at dawn. 
We empirically adopt a threshold of 1.3 times of the median value of the errors and exclude those data with error larger than the threshold. The good data ratio of each light curve are listed in the column 7 of Table \ref{observation}.

\subsection{Error scaling}\label{err_scale}
The photometric error yielded by APPHOT in IRAF is believed to be underestimated. Hence, a factor $\eta$ is introduced to amplify the underestimated error. Different values were calculated by different authors e.g. 1.3 by \citet{2005MNRAS.358..774B}, 1.5 by \citet{2004MNRAS.350..175S}, \citet{2003ApJ...586L..25G} and 1.73 by \citet{1999MNRAS.309..803G}, etc. 

\begin{figure}
\includegraphics[width=\columnwidth]{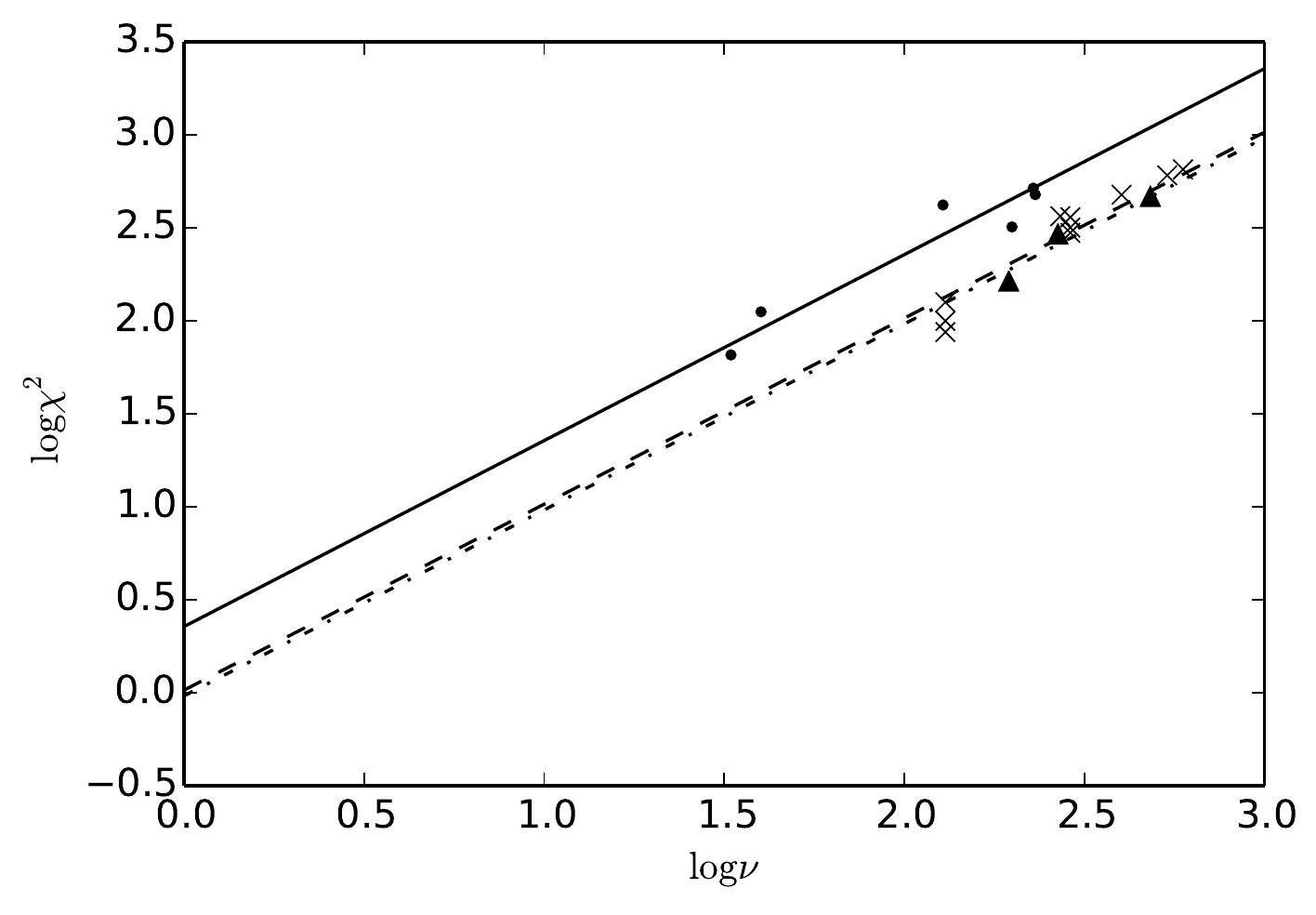}
\caption{
    Regression with $\log\chi^2$ and $\log\nu$ of each telescope. The crosses, dots and triangles represent data of the 85 cm, 80 cm and 60 cm telescopes, respectively. The dashed, solid and dash-dotted lines represent the fitting curves of the 85 cm, 80 cm and 60 cm telescopes, respectively.}
\label{eta}
\end{figure}

Before subsequent analysis, we should estimate the scale factor $\eta$ for each telescope. The method of \citet{2013JApA...34..273G} is adopted. First, for each light curve, we calculate the $\chi^2$ value of differential magnitudes of the check star by using the equation as follow:
\begin{equation}\label{chi2}
\chi^{2}=\displaystyle\sum_{i=1}^{N}\frac{(V_i-\overline{V})^2}{\sigma_{i}^2} ,
\end{equation}
where $V_i$ is the $i$th differential magnitude, $\overline{V}$ is the mean of all differential magnitudes, and $\sigma_i$ is the error of $V_i$, which is propagated from raw photometry errors of stars. The corresponding degree of freedom 
\begin{equation}
\nu=N-1=\displaystyle\sum_{i=1}^{N}\frac{(V_i-\overline{V})^2}{\eta^2\sigma_{i}^2}=\chi^2/\eta^2.
\end{equation}
Then we perform a regression analysis with $\log\chi^2$ and $\log\nu$ with a fixed slope 1. The intercept $K$ are obtained from 
\begin{equation}
\log\chi^2=K+\log\nu,
\end{equation}
where $10^{K}=\eta^2$. We obtain $\eta=1.02$ for the 85 cm telescope, $\eta=1.51$ for the 80 cm telescope and $\eta=0.98$ for the 60 cm telescope. The fitting lines are plotted in Figure \ref{eta}. Because there's only one light curve of the 2.16 m telescope and the 85 cm telescope with old CCD, we use reduced $\chi^2$ values to represent the $\eta^2$. The $\eta$s of the 2.16 m telescope and the 85 cm telescope with old CCD are 1.28 and 1.20, respectively. In the next section, our analysis are based on the error-scaled data.

It is clear that the scale factors of the 85 cm telescope and the 60 cm telescope are much less than that of the 80 cm telescope. This offset may be caused by some unknown interferences in the 80 cm telescope light path, which adds extra fluctuations after flat-fielding and amplifies the $\eta$ factor. Also, this unknown interference could create pseudo flares in the light curves of the object. 

\begin{figure*}
\centering
\includegraphics[width=1.99\columnwidth]{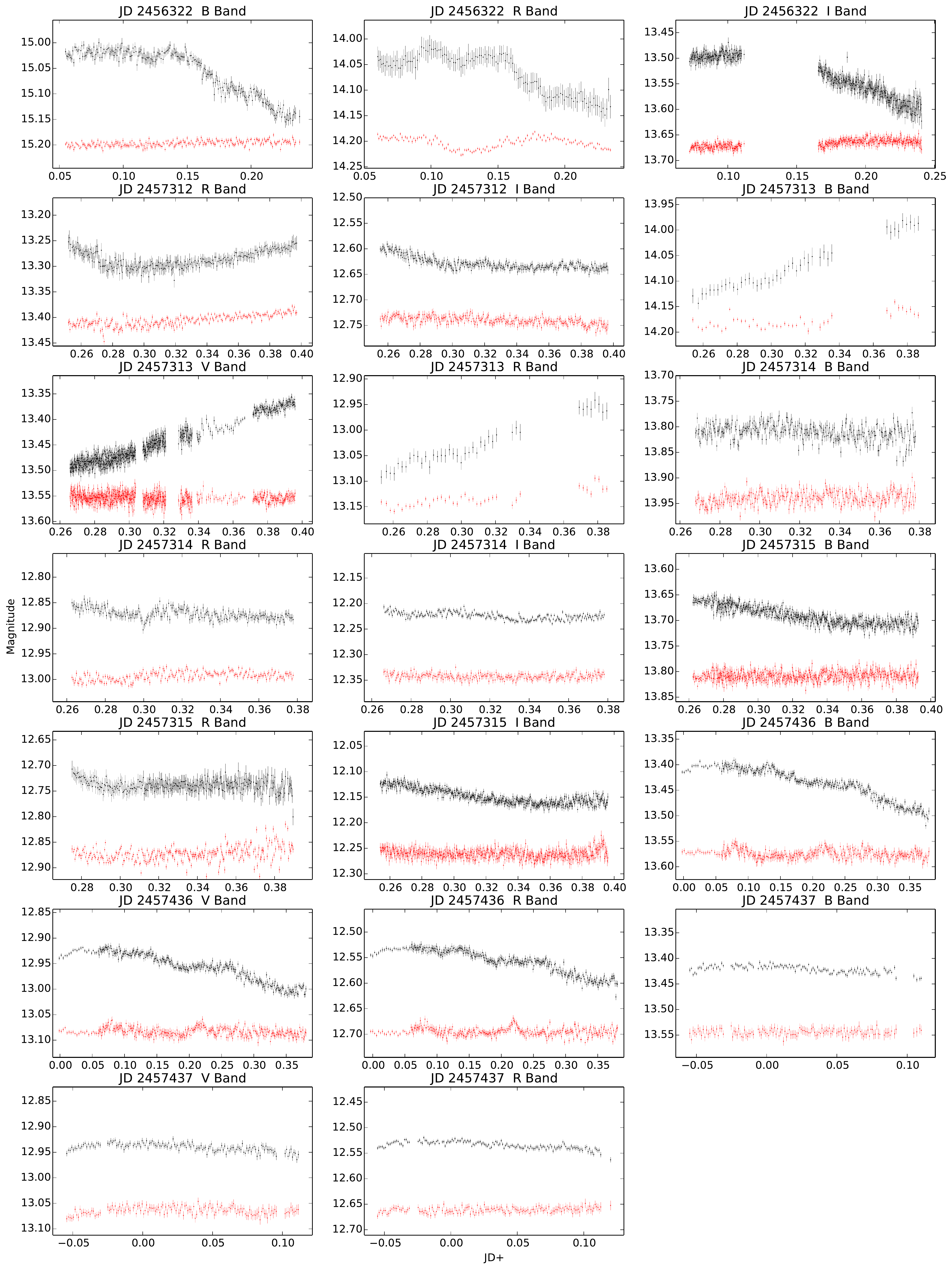}
\caption{Intra-day light curves in each band on each day. The light curves of the check star are plotted in red dots and shifted to a proper position under the source's light curves.}
\label{individual_lc}
\end{figure*}

\begin{figure}
\centering
\includegraphics[width=0.99\columnwidth]{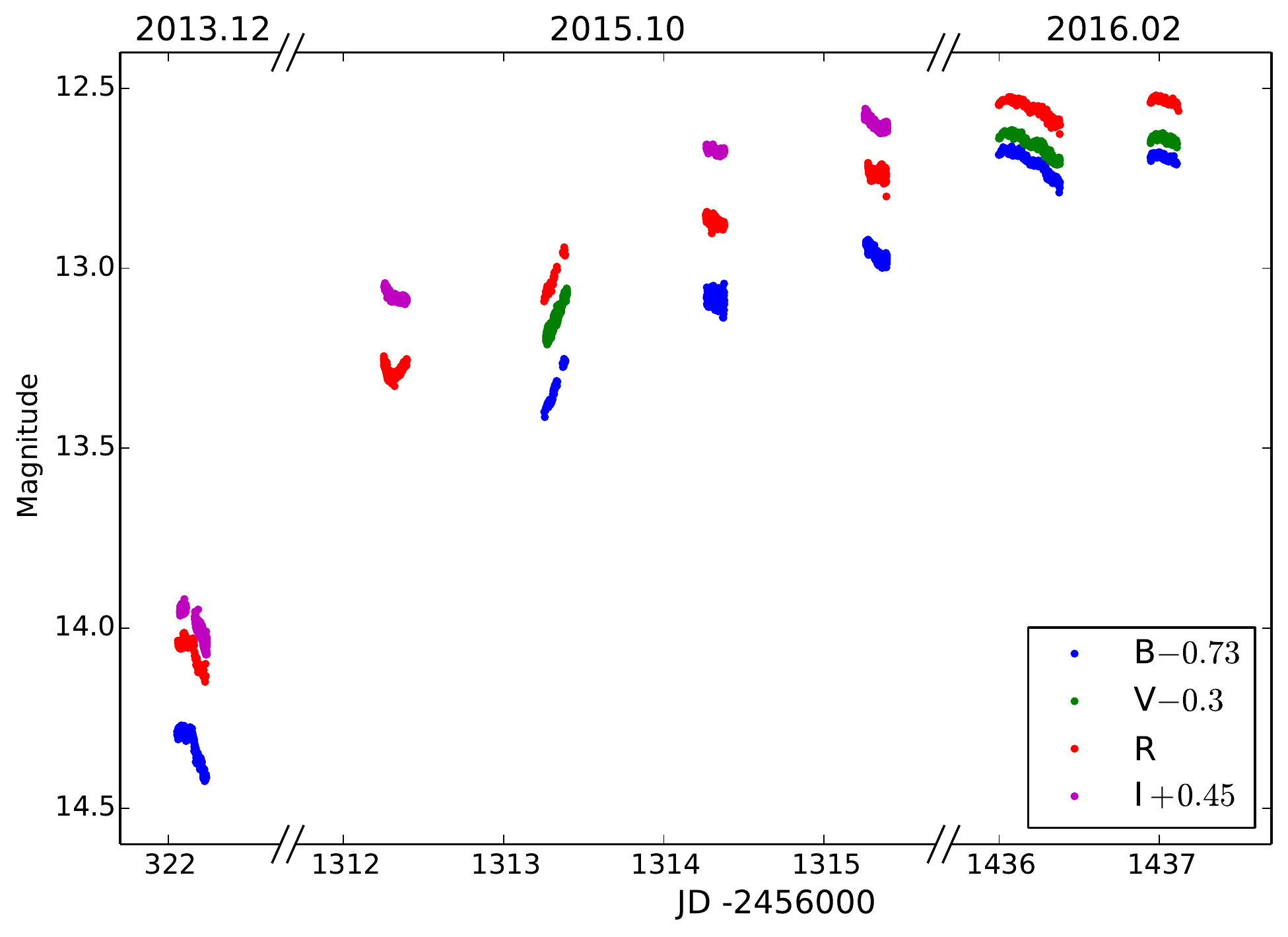}
\caption{Overall multi-band light curves in seven days. Light curves in the $B$, $V$ and $I$ bands are shifted.}
\label{overall_lc}
\end{figure}

\section{Results}\label{results}
\subsection{Light curves}\label{sec_lc}
The intra-day light curves of the source are plotted in Figure \ref{individual_lc}. The differential light curves of the check star is also plotted to indicate the photometry accuracy. The overall light curves in seven days are plotted in Figure \ref{overall_lc} to show the inter-day variability.  During our observation, the source was faintest on JD 2456322 with 14.15 mags in the $R$ band and reached the brightest state on JD 2457437 with 12.51 mags in the $R$ band.

It is clear that the variations on JDs 2456322, 2457313 and 2457436 are significant. On JD 2456322, variation in the $B$ band reaches 0.15 mags, and the maximum variation rate is 0.085 mags h$^{-1}$. On JD 2457313, the source turned bright monotonously in all three bands with a variation rate of about 0.053 mags h$^{-1}$. On JD 2457436, the light curves descend with a moderate rate and a plateau can be seen in all bands. On JDs 2457314 and 2457437, the source shows marginal intra-day variations in all bands. On JD 2457315, the brightness descended and then rose again in the $B$ and $I$ bands. Inter-day variations are considerable. From JDs 2457312 to 2457315, the R band magnitude varied 0.56 mags within 4 days.

We notice that the check star differential light curves observed by the 80 cm telescope are slightly unstable. This phenomenon are consistent with the large $\eta$ value obtained in Section \ref{err_scale}. The unknown interference in the 80 cm telescope affects the photometry accuracy. When the source has significant IDV, the influence tends to be relative low, e.g. light curves on JDs 2456322 and 2457313. However, when the source is only marginally variable, this influence could alter the entire profile of the light curve. On JDs 2457312,  2457314 and 2457315, light curves in the $R$ band are considerably different from that in other bands. The incongruity could produce pseudo results in colour variation and cross-correlation analyses.
As a result, the $R$ band light curves on that three days should be ruled out from these two analyses.

\subsection{IDV test}
To quantitatively test the intraday variability of the source, we adopt two up-to-date robust statistical tests. They are enhanced $F$-test and nested analysis of variance (ANOVA) \citep{2015AJ....150...44D}.

In the original $F$-test \citep{2010AJ....139.1269D}, the statistical value $F$ is obtained from the variances of the object's and a check star's differential magnitudes. The enhanced $F$-test includes several field stars' data to produce a more robust result. In this test, we use the brightest unsaturated star 2 as comparison star to get the differential magnitudes of the object and field stars 3 to 8. For stars 3 to 8, we perform following procedures to get the stacked differential light curves. First, we fit the mean differential magnitudes and the corresponding standard deviations with an exponential curve. Then we scale the variance of each differential light curve to the same level of the object. Finally we subtract the mean value of each scaled differential light curve. The variance of the stacked differential light curve of field stars are calculated as the denominator of the $F$ value. The $F$ value, two corresponding degrees of freedom and the probability to pass the null hypothesis are listed in columns 3 to 6 in Table \ref{test}.

In the nested ANOVA, multiple field stars are involved as comparison stars. In our test, stars 2 to 8 are used to calculated the differential light curves of the object. Then the 7 differential light curves are divided into a number of groups with 5 points in each group. We follow  Equation 4 in \citet{2015AJ....150...44D} to test the null hypothesis that the deviation of the mean values of differential light curves in each group is zero. The statistical value $F$, two degrees of freedom and the probability to pass the null hypothesis are listed in columns 7 to 10 in Table \ref{test}. Meanwhile, in order to test the invariability of the check star, we also apply the test to star 5 and present the results in columns 11 to 14.

For most observation sessions, the light curves of the object pass both tests and the light curves of the check star fail to pass the nested ANOVA. So the object was variable in these sessions.
In two observation session of the 80 cm telescope, the light curves of the object fail to pass the $F$ test, and the light curves of the check star pass the nested ANOVA. This kind of behaviour is caused by the unknown interference in the light path of the 80 cm telescope. The light curve of each object in the field are added with pseudo variations. We cannot determine the variability in these sessions. In another two sessions of the 80 cm telescope, according to the same reason, the light curves of the check star pass the nested ANOVA, but the object's light curves pass both tests as well. We regard that the object was variable. All abnormal behaviours observed by the 80 cm telescope are marked with $*$.
In four sessions of other telescopes, check star's light curves pass the nested ANOVA and are marked with $**$. We ascribe them to the the marginal variations of the check star. Considering that the light curves of the object pass both tests, we regard that the object was  variable. In additional to these abnormal events, on JD 2457314, the light curve in the $B$ band fail to pass the enhanced $F$ test but pass the nested ANOVA, indicating that the IDV is only beyond the limit of photometry accuracy. The object was probably variable on that night. 
In conclusion, the object was variable in all seven nights.

\begin{table*}
\centering
\caption{Results of IDV test.}
\label{test}
\begin{tabular}{ccrrrcrrrcrrrcc}
\hline
Julian Date & Filter & \multicolumn{4}{c}{Enhanced-F test} & \multicolumn{4}{c}{Nested ANOVA for blazar} & \multicolumn{4}{c}{Nested ANOVA for Star 5} & Variable \\
  & & $F$ &$\nu_1$& $\nu_2$ & $P$ & $F$ & $\nu_1$ & $\nu_2$ & $P$ & $F$ & $\nu_1$ & $\nu_2$ & $P$\\
\hline
2456322&$B$&29.06&229&1374&$<0.0001$&61.81&45&184&$<0.0001$&1.06&45&184&0.3830&V\\
              &$I$&16.83&423&2538&$<0.0001$&74.64&83&336&$<0.0001$&6.17&83&336&$<0.0001^{**}$&V\\
              &$R$&9.75&128&768&$<0.0001$&152.59&24&100&$<0.0001$&23.49&24&100&$<0.0001^{*}$&V\\
2457312&$I$&2.97&266&1596&$<0.0001$&18.94&52&212&$<0.0001$&1.81&52&212&0.0018&V\\
              &$R$&1.84&228&1368&$<0.0001$&20.93&44&180&$<0.0001$&6.55&44&180&$<0.0001^{*}$&V\\
2457313&$B$&13.96&43&258&$<0.0001$&102.30&7&32&$<0.0001$&4.26&7&32&0.0020&V\\
              &$R$&9.80&38&228&$<0.0001$&28.02&6&28&$<0.0001$&2.82&6&28&0.0029&V\\
              &$V$&13.28&616&3696&$<0.0001$&98.74&122&492&$<0.0001$&1.09&122&492&0.2557&V\\
2457314&$B$&1.16&271&1626&$0.0512$&2.22&53&216&$<0.0001$&1.33&53&216&0.0838&P\\
              &$I$&1.71&212&1272&$<0.0001$&5.77&41&168&$<0.0001$&1.14&41&168&0.2815&V\\
              &$R$&0.79&199&1194&$0.9818^{*}$&15.97&39&160&$<0.0001$&4.46&39&160&$<0.0001^{*}$& \\
2457315&$B$&2.89&476&2856&$<0.0001$&17.65&94&380&$<0.0001$&0.77&94&380&0.9352&V\\
              &$I$&5.45&482&2892&$<0.0001$&26.66&95&384&$<0.0001$&1.16&95&384&0.1708&V\\
              &$R$&0.63&231&1386&$1.0000^{*}$&10.48&45&184&$<0.0001$&4.31&45&184&$<0.0001^{*}$& \\
2457436&$B$&22.41&289&1734&$<0.0001$&179.57&57&232&$<0.0001$&4.72&57&232&$<0.0001^{**}$&V\\
              &$R$&7.18&290&1740&$<0.0001$&121.56&57&232&$<0.0001$&3.11&57&232&$<0.0001^{**}$&V\\
              &$V$&12.26&286&1716&$<0.0001$&111.60&56&228&$<0.0001$&2.53&56&228&$<0.0001^{**}$&V\\
2457437&$B$&1.89&130&780&$<0.0001$&12.00&25&104&$<0.0001$&0.97&25&104&0.5106&V\\
              &$R$&1.60&133&798&$<0.0001$&8.27&25&104&$<0.0001$&1.28&25&104&0.1966&V\\
              &$V$&1.87&152&912&$<0.0001$&9.29&29&120&$<0.0001$&1.74&29&120&0.0202 &V\\
\hline
\multicolumn{14}{l}{$^*$ Caused by the unknown interference in the light path of the 80 cm telescope.  }\\
\multicolumn{14}{l}{$^{**}$ Caused by the marginal variation of the check star.  }
\end{tabular}
\end{table*}

\subsection{Colour variations}
\begin{figure*}
\includegraphics[width=1.89\columnwidth]{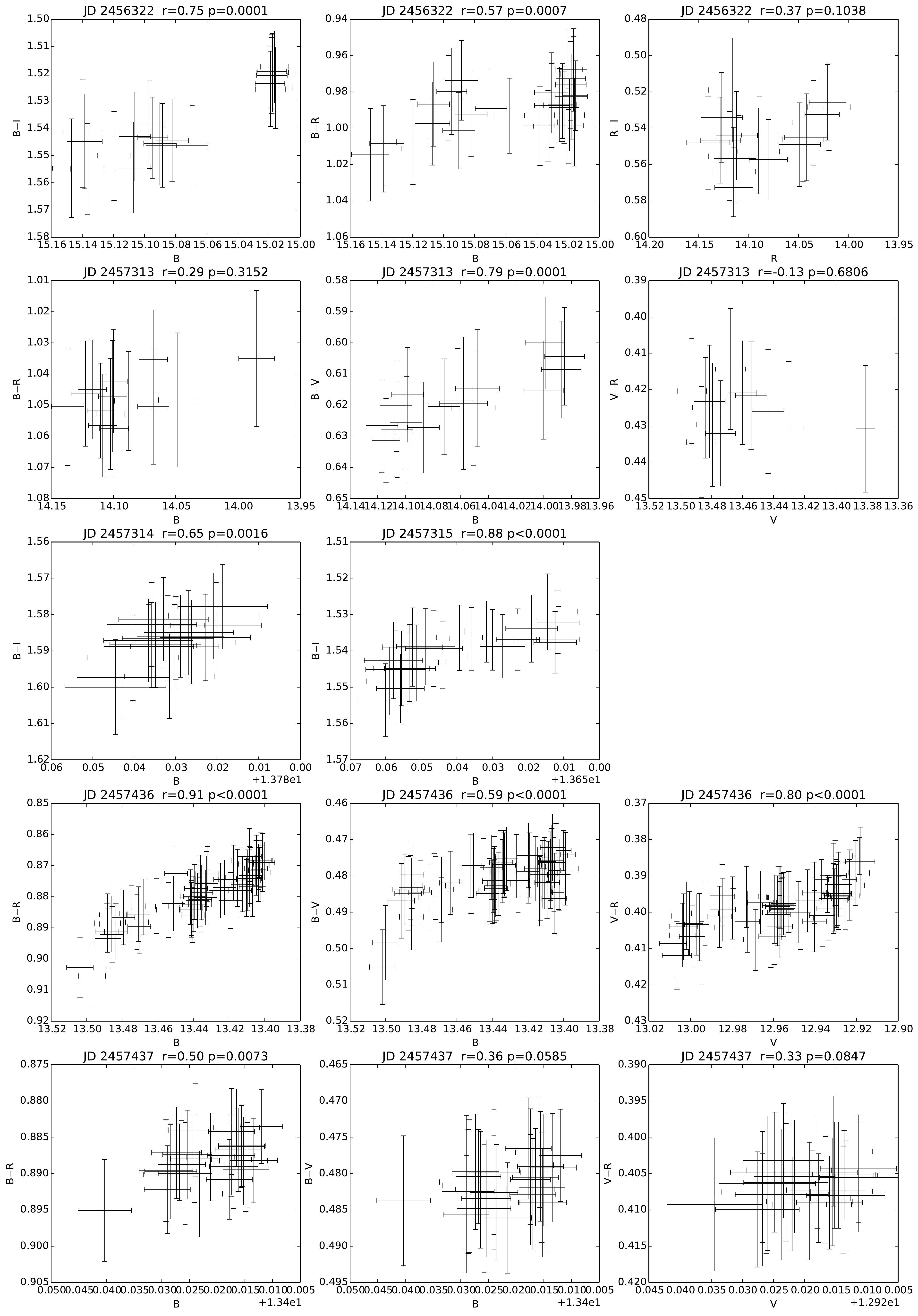}
\caption{Colour-magnitude diagrams of 8-minute binned light curves.}
\label{colour_idv}
\end{figure*}

To investigate the relationship between spectral changes and flux variations, we first calculate 8-minute binned light curves, colours, and then plot the colour-magnitude diagrams. Results are shown in Figure \ref{colour_idv}. The $R$ band light curve on JDs 2457312, 2457314 and 2457315 are excluded. We also calculate Spearman correlation coefficient and the associated  p-value for each diagram.  On JD 2456322, the p-value of $B-I$ versus $I$ on JD 2456322 reaches 0.0001, indicating a mild bluer-when-brighter (BWB) colour behaviour. On JD 2457436, p-values of three diagrams are less than 0.0001, indicating a significant BWB trend. On JD 2457313, the source shows almost no colour variation. Though the p-value of $B-V$ reaches 0.0001, it could be caused by the slight systematic bias between the 85 cm and the 80 cm telescopes, because the colour index $B-R$ shows no variability. On JDs 2457314 and 2457437, the magnitude variations are too weak to show any colour variation. On JD 2457315, the source showed an overall BWB trend but turned to be achromatic at the bright end. The p-value confirms the colour variation.

\begin{figure}
\begin{tabular}{c}
\includegraphics[width=.99\columnwidth]{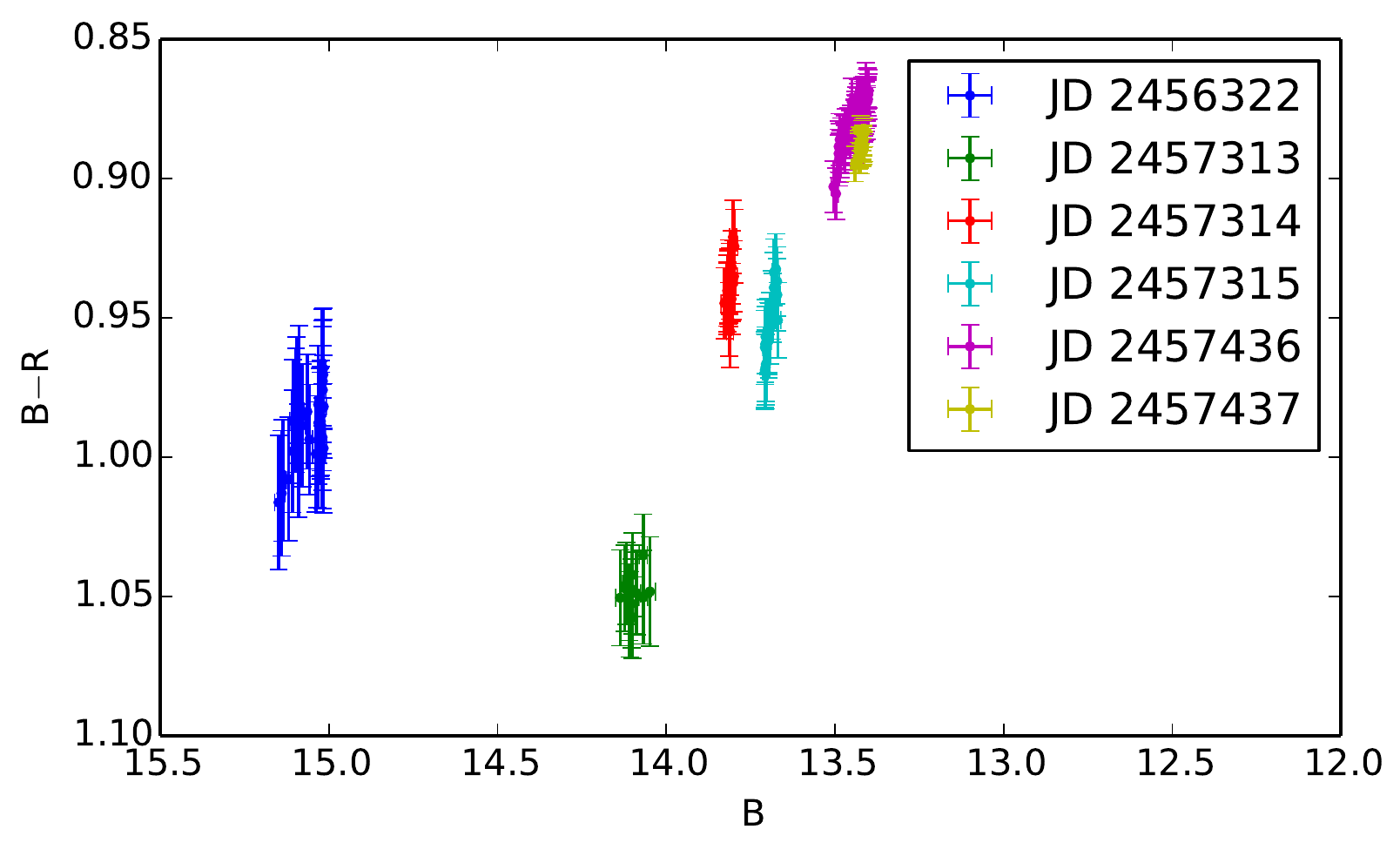}\\
\end{tabular}
\caption{Long term colour-magnitude diagram between the $B$ and $R$ bands. Different colour indicate colour behaviours in different days.}
\label{colour_long}
\end{figure}

We also plot the overall colour-magnitude diagram of the $B$ and $R$ bands in Figure \ref{colour_long}. The interday colour behaviour doesn't follow an overall BWB trend. Colour index $B-R$ shifted $\sim0.1$ mags from JDs 2457313 to 2457314. Though the $R$ band data of 2013 and 2015 are obtained by the 80  cm telescope with additional uncertainties,  the 0.1 mags inter-day colour variation is far beyond the instrumental uncertainties. If we take the average value of colour index from JDs 2457313 to 2457315, the long-term colour behaviour follows a mild BWB trend. 

For different blazars, two distinct behaviours have been observed, i.e. redder-when-brighter (RWB) and BWB. The RWB behaviour appears mostly in FSRQs, while BL Lac objects exhibit more BWB events. On the timescale from months to years, for example, two of three FSRQs show RWB and three of four BL Lac objects show BWB in \citet{2006A&A...450...39G}'s work. Four of six FSRQs are RWB and three of six BL Lac objects are BWB reported by \citet{2010MNRAS.404.1992R}. Eight of nine FSRQs are RWB in SMARTS campaign \citep{2012ApJ...756...13B}. In the monitoring by \citet{2011PASJ...63..639I}, three of seven FSRQs are RWB when they are in faint states while twenty-three of 27 BL Lac objects exhibit BWB behaviour.

Most reported BWB chromatism of S5 0716+714 is on short timescale \citep{1997A&A...327...61G,2003A&A...402..151R,2012AJ....143..108W,2013ApJS..204...22D,2014MNRAS.443.2940H}. However, \citet{2006MNRAS.366.1337S} and \citet{2009ApJS..185..511P} found no clear evidence of intra-day BWB behaviour. Our intraday results consist with most historical observations. On the long timescale, this BWB chromatism turns meagre, i.e. the long term colour index versus magnitude slope is smaller than short term one (e.g. \citealt{2007AJ....133.1599W}). Our overall colour behaviour show this trend as well. \citet{2003A&A...402..151R} reported that the source tends to be achromatic on a ten-year timescale. Our results are broadly consistent with the historical observation data.

\subsection{Cross-correlation analysis}
To investigate the inter-band time lags, we perform cross-correlation analysis. There are a couple of mathematical cross-correlation functions (CCF) and methods to detect lags and estimate the uncertainties. We adopt two sets of them. 

The z-transformed discrete correlation function (ZDCF), introduced by \citet{1997ASSL..218..163A}, corrects several biases of the discrete correlation function (DCF) \citep{1988ApJ...333..646E} by using equal population binning and Fisher's z-transform. \citet{2013arXiv1302.1508A} offered a Fortran program PLIKE to calculate the ZDCF peak location based on the maximum likelihood (ML) estimation. The likelihood value of the $i$th point on ZDCF curve, $L_i$, is approximately the probability that the correlation coefficient of this point is larger than that of any other ZDCF points. Hence, the maximum likelihood value is at the highest ZDCF point. The uncertainty of the lag, defined as the 68.2\% fiducial interval of the normalised likelihood function.

Another set of methods is proposed by \citet{1998PASP..110..660P,2004ApJ...613..682P}, which employs a Monte-Carlo (MC) method to estimate the peak position $\tau_{\mathrm{peak}}$ or the centroid position $\tau_{\mathrm{cent}}$ of an interpolated cross-correlation function (ICCF) and their uncertainties. In each MC realisation, both "flux randomisation" (FR) and "random subset selection" (RSS) processes are applied. 
The peak position $\tau_{\mathrm{peak}}$ and intensity $r_{\mathrm{max}}$ are obtained directly from the ICCF curve, and $\tau_{\mathrm{cent}}$ is calculated by points above a threshold, which is typically $0.8 r_{\mathrm{max}}$. After a large number of MC realisations, distributions of cross-correlation centroid (CCCD) and cross-correlation peak (CCPD) are built. The value of $\tau_{\mathrm{centroid}}$/$\tau_{\mathrm{peak}}$ and its uncertainty are derived from the mean and the 1$\sigma$ deviation of CCCD/CCPD. If the CCCD/CCPD is asymmetric, the lower and upper uncertainties are defined as values that correspond to 15.87\% and 84.13\% of the cumulative distribution function of CCCD/CCPD, respectively. For a broad ICCF peak, \citet{1998PASP..110..660P} recommended using $\tau_{\mathrm{cent}}$ rather than $\tau_{\mathrm{peak}}$.

\begin{figure*}
\includegraphics[width=1.99\columnwidth]{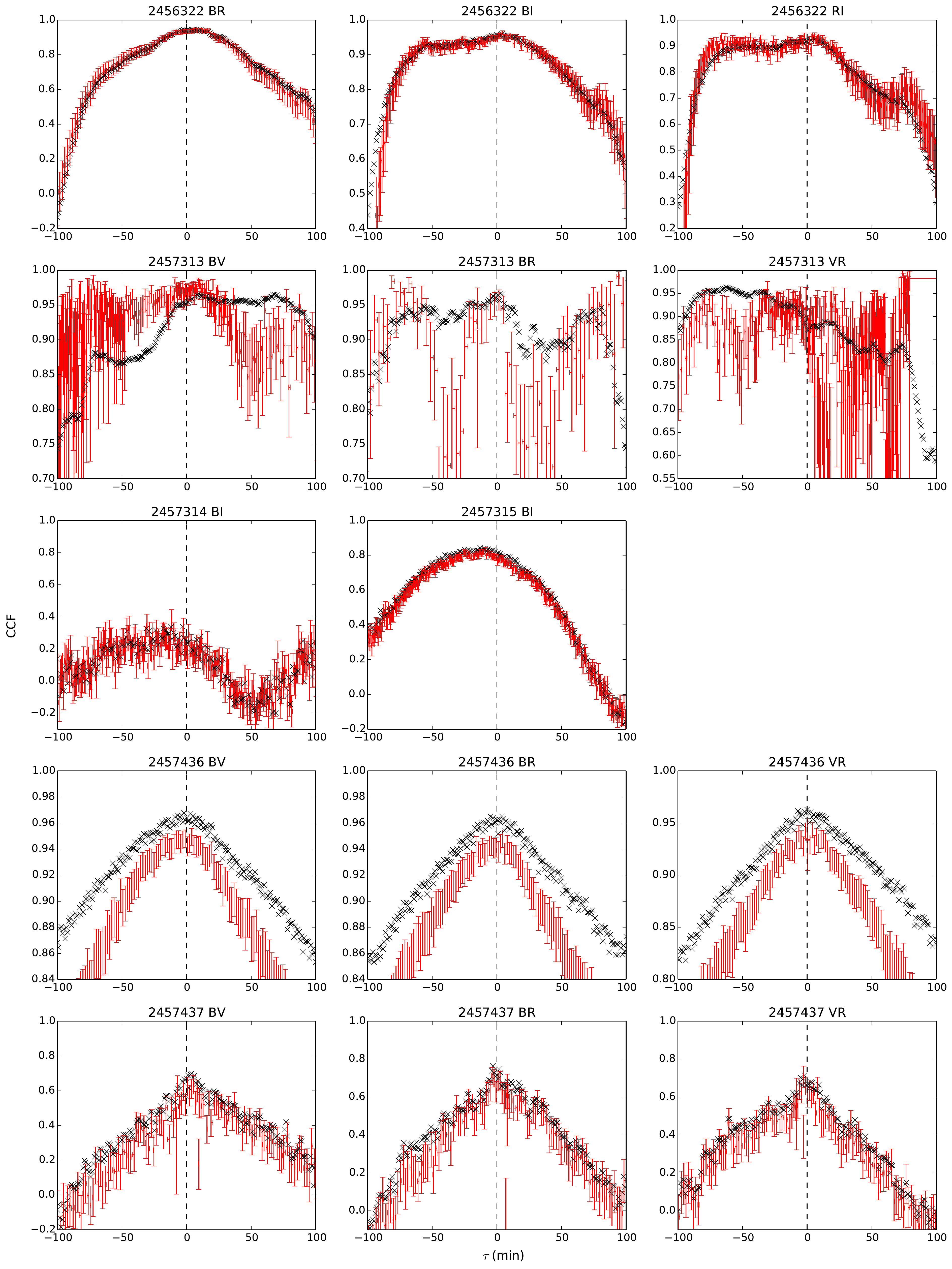}
\caption{Results of cross-correlation analysis. The black and red points are ICCF and ZDCF curves respectively. The black dashed lines indicate zero-lags.}
\label{ccf}
\end{figure*}

The ZDCF and ICCF curves are plotted in Figure \ref{ccf}. The $R$ band light curves on JDs 2457412, 2457414 and 2457415 are excluded for cross-correlation analysis. On JD 2456322, curves of two CCFs are most the same. Because of the the vacancy interval on the $I$ band light curve,  the left side of peaks of $B$ versus $I$ and $R$ versus $I$ are flat. We adjust the threshold for centroid position estimation to a higher level properly. 
On JD 2457313, significant deviations between the ZDCF and ICCF curves can be seen. Due to bad weather condition at that night, we clip data points with large photometric error (see Section \ref{reduction}). The refined light curves have several significant gaps. Because ZDCF and ICCF have different strategies of sample collecting for correlation calculation, these unevenly sampled light curves could cause the deviations.
On JD 2457314, the low variation amplitude makes the CCF curves with low correlation coefficient values, which is lower than the threshold for FR/RSS method. On JDs 2457315, 2457436 and 2457437, CCF curves show significant symmetric peak.

We perform ML estimation for the ZDCF peak position and five thousands FR/RSS processes for the ICCF peak centroid estimation. 
Results of estimated time lags of two methods as well as their CCF values at peak positions are listed in Table \ref{lags}. 
On JD 2456322, positive results estimated by ZDCF/ML method indicates that variation in the $B$ band leads that of the $R$ and $I$ bands, but no lag is detected by using the FR/RSS method. 
On JD 2457315, variations in the $B$ band lagged that in the $I$ band by $8.7\substack{+6.9\\-2.1}$ minutes by ZDCF/ML method and by $16.5\pm3.1$ minutes by FR/RSS method. The zero point is out of the 1 $\sigma$ intervals of two methods, and even out of the 3 $\sigma$ confidence interval calculated by ICCF+FR/RSS method, which is 6.1 to 25.9 minutes. The estimated lags by two methods are somewhat different, which is caused by the different estimation approaches. The ZDCF-ML method only considers the position of the maximum CCF value as the lag, if the maximum CCF point is not at the centroid, the fiducial distribution of likelihood can be skew, which leads to an asymmetric uncertainty interval. On JD 2457315, the maximum ZDCF point is at the right side of the centroid. For this kind of broad CCF peaks, the FR/RSS method for centroid estimation is more appropriate. The ZDCF/ML method, though have some deviation, proves the lag and constrains the lag value.

\begin{figure}
\begin{tabular}{c}
\includegraphics[width=.99\columnwidth]{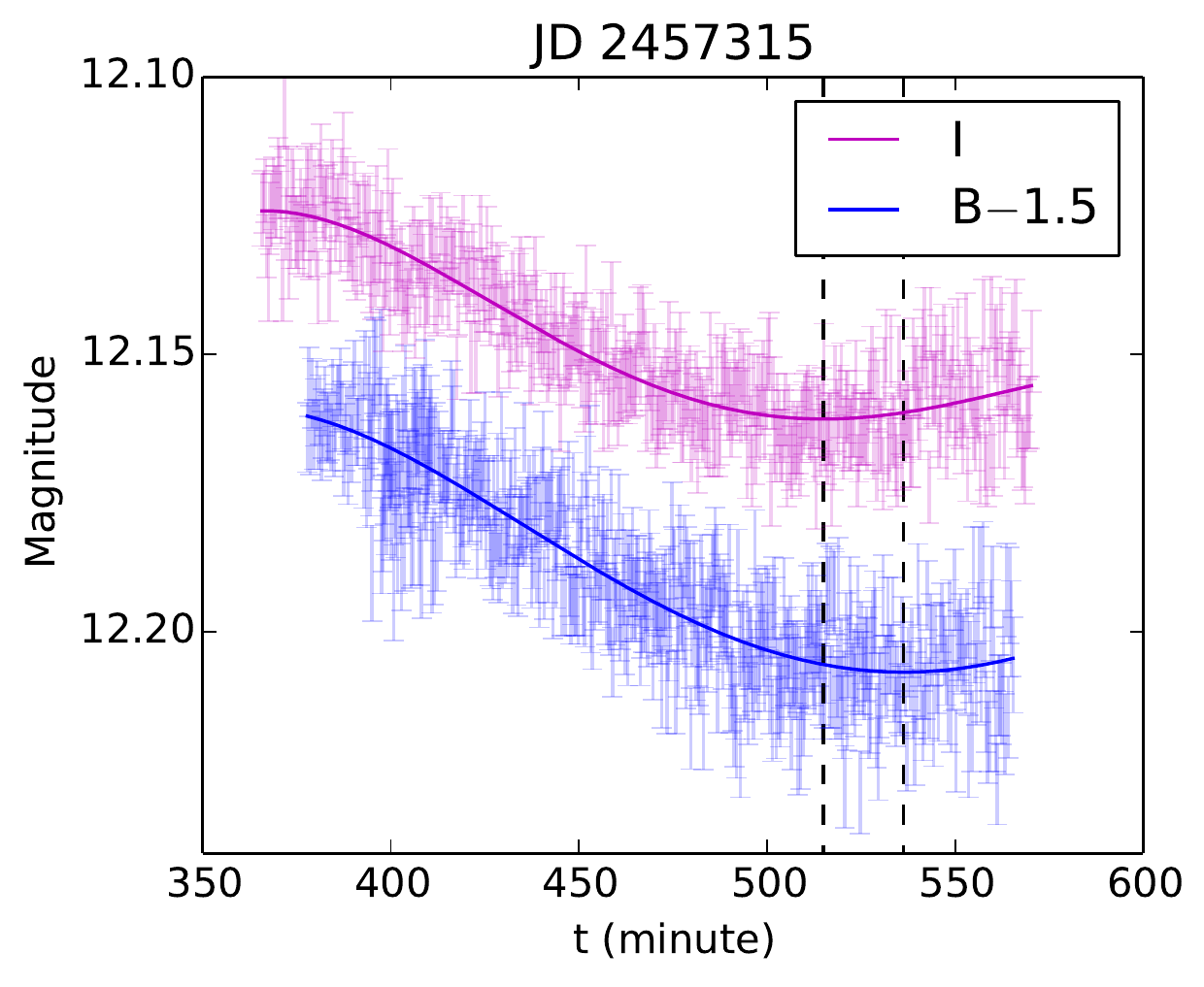}\\
\includegraphics[width=.99\columnwidth]{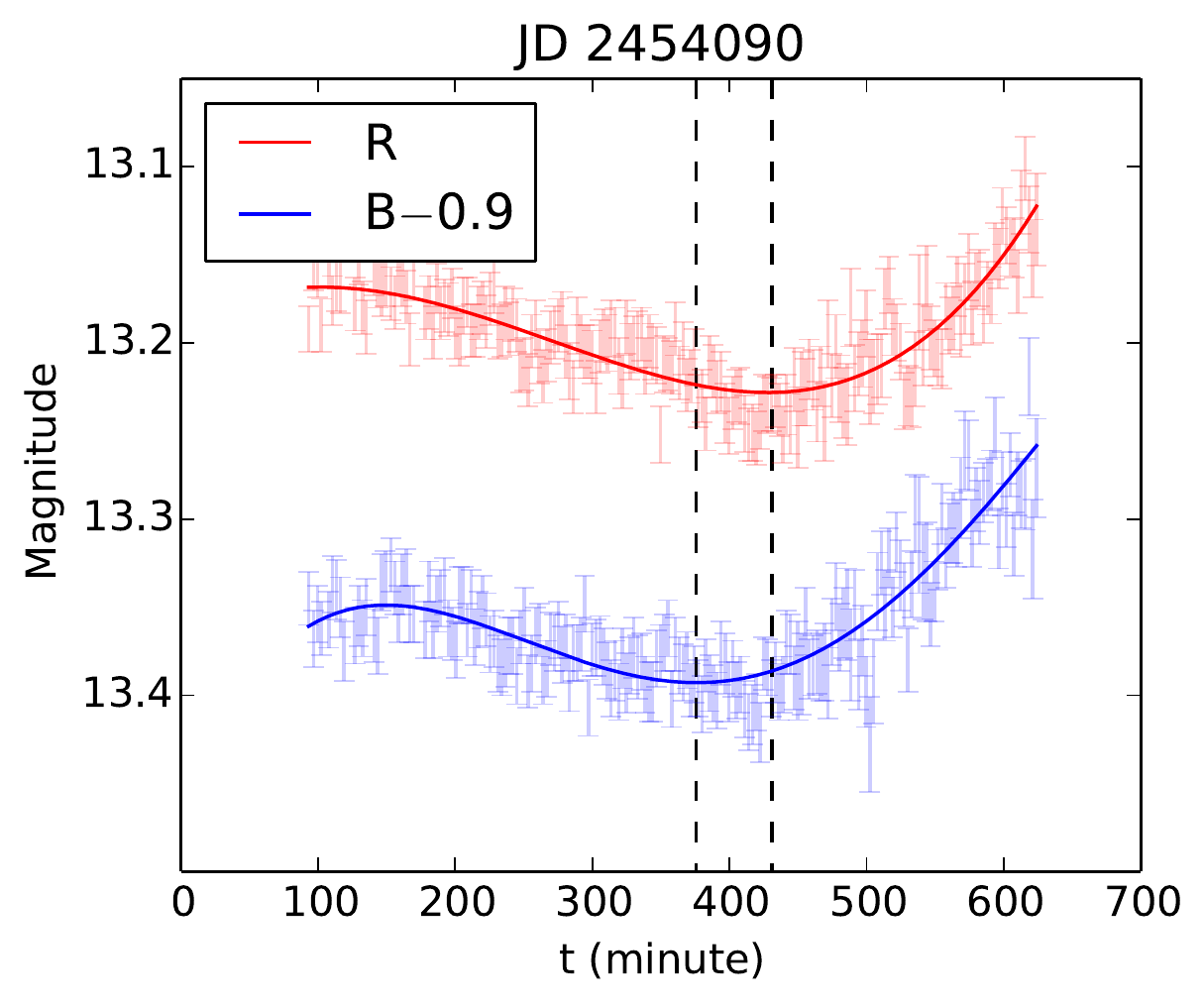}
\end{tabular}
\caption{Top panel: polynomial fittings for the light curves in the $B$ and $I$ bands on JD 2457315. The $B$ band light curve is shifted. Black dashed lines are positions of minima in two light curves. Lower panel: the same as the top panel for the light curves on JD 2454090 observed by Wu et al. 2012.}
\label{fitting}
\end{figure}

We perform fourth-ordered polynomial fittings for the light curves and calculate the minimum positions, see the upper panel of Figure \ref{fitting}. The $\sim20$ mins lag between two minima is close to the results of the cross-correlation. We compare this result with the lag on JD 2454090 observed by \citet[hereafter Wu12]{2012AJ....143..108W} (the lower panel of Figure \ref{fitting}). There are a couple of similarities between these two lags: the values of delays are both of tens of minutes; they are both at the junctions of two flares; at least one flare follows the BWB chromatism. The difference is that the variation in the long wavelength leads that at the short wavelength on JD 2457315, while the result on JD 2454090 is the reverse.

\begin{table}
\centering

\renewcommand{\arraystretch}{1.5}
\caption{Results of cross-correlation analysis. Positive values indicate that the previous band leads the latter band.}
\label{lags}
\begin{tabular}{ccrcrc}

\hline
JD& Passbands & \multicolumn{2}{c}{ZDCF-ML} & \multicolumn{2}{c}{ICCF-FR/RSS} \\
&&Lag (min)&ZDCF$_\mathrm{peak}$&Lag (min)&ICCF$_{\mathrm{peak}}$\\
\hline
2456322 & $B-R$ & $10.6\substack{+4.8\\-7.5}$ &0.94&  $4.0\pm5.8$ &0.94\\
& $B-I$ & $4.6\substack{+5.2\\-2.5}$ & 0.96  & $1.3\pm6.9$ & 0.96\\
& $R-I$ & $5.9\substack{+3.7\\-40.1}$ &0.94& $-7.9\pm18.0$&0.93\\
2457313 & $B-V$ & $-8.9\substack{+18.9\\-4.1}$ &0.98& $-5.5\pm9.6$&0.96\\
& $B-R$ & $1.6\substack{+2.5\\-9.2}$ &0.96& $-2.7\pm36.8$ &0.97\\
& $V-R$ & $-30.5\substack{+24.8\\-2.3}$ &0.96& $-9.7\pm17.7$&0.96\\
2457314 & $B-I$ & $-11.6\substack{+8.7\\-27.8}$ &0.38& \multicolumn{1}{c}{-}&0.32\\
2457315 & $B-I$ & $-8.7\substack{+2.1\\-6.9}$ &0.83& $-16.5\pm3.1$&0.84\\
2347436 & $B-V$ & $4.0\substack{+4.6\\-11.8}$ &0.95&  $-2.9\pm5.4$&0.97 \\
& $B-R$ & $4.5\substack{+3.6\\-10.6}$ &0.94&  $-4.3\pm6.0$&0.97\\
& $V-R$ & $0.5\substack{+6.8\\-4.2}$ &0.94& $-0.2\pm5.2$&0.96\\
2457437 & $B-V$ & $5.2\substack{+3.6\\-7.9}$ &0.64& $6.9\pm13.9$&0.70\\
& $B-R$ & $-3.7\substack{+5.4\\-1.1}$ &0.71&  $-0.9\pm1.8$&0.76\\
& $V-R$ & $-2.7\substack{+8.3\\-1.9}$ &0.71&  $-3.4\pm13.1$&0.68\\
\hline
\end{tabular}
\end{table}

\section{Discussion}\label{discussion}
During our multi-band observation, mild BWB colour behaviours and one inter-band time lag were observed.

Several models can interpret the BWB chromatism. Within a
one-component synchrotron model, a flatter spectrum or a lower spectral index
at the high state indicates a flatter relativistic electron spectrum. When the
source gets brighter, more electrons are accelerated by the shock and injected
into the emission region \citep{2004A&A...419...25F}. The BWB trend can be 
interpreted by a two-component model as well, which includes an underlying component
at the red side and a broadband achromatic variable synchrotron component,
especially when the source is faint \citep{2015A&A...573A..69W}. This
underlying component could be from the host galaxy, whose SED peak locates at
near-infrared frequencies. But for S5 0716+714, the magnitude of the host
galaxy is only 17.5 in the $I$ band \citep{2008A&A...487L..29N}. Such a faint host galaxy only
contributes less than 0.02 mags of $B-I$ colour variation when the source
turns bright from 13.5 to 13.3 mags in the $I$ band. This contribution could
somehow lead to a marginal BWB trend at a dim state, as in the case of the $B-I$ colour behaviour on 
JD 2456322, but cannot explain significant colour
variabilities at high states (e.g. \citealt{2005AJ....129.1818W}). Another
interpretation is synchrotron peak shift. Since the synchrotron peak of S5 0716+714 locates
at near IR to UV frequencies, optical colour index should be sensitive to the
peak shifts. \citet{2014ApJ...783...83L} reported that the
$\nu_{\mathrm{syn\_peak}}$ exceeded beyond the $V$ band when the source was in
a bright state. The peak shift could be caused either by
high energy electron injection or by variation of the Doppler factor $\delta$
\citep{2011PASJ...63..639I}. However the factor of $\delta$ variation might be ruled out,  
because the Doppler factor amplifies the emission coefficient by three orders of magnitude but only one
for the observational frequency. \citet{2003A&A...402..151R} explains the long term variation
of this source by a variation of $\delta\sim1.3$. Such a subtle change only 
accounts for achromatic behaviours. 

Inter-band time lags indicate inconformities of variations at different wavelengths, or the so-called spectral hysteresis. The previous spectral hysteresis study focuses on the whole profile of flares (e.g. \citealt{2000ApJ...541..153F,1999ApJ...527..719Z}). Injection and acceleration of relativistic electrons into the radiation zone and subsequent radiative cooling process can account for the observed phenomena \citep{1998A&A...333..452K}. 
However, we only observed a time lag between the minima of two light curves in different bands.
We propose a possibility to produce such lags at trough positions. The left part of the bottom on JD 2457315  and the right part on JD 2454090 are mild BWB with the opposition parts nearly achromatic. The nearly achromatic variation changes the time scales of previous/subsequent BWB variation at different wavelengths. Variation in the long-wavelength band has shorter time scale than the short-wavelength band's. This could explain why the variations in the long-wavelength band lead that in the short-wavelength band on JD 2457315, but lagged on JD 2454090. 

In general, this kind of ``lags'' could appear at troughs where the adjacent flares have different colour behaviours. It raises another question, why we only observed a small number of them? Four main parameters which determine the lag detection have been discussed by \citet{2012AJ....143..108W}, they are wavelength separation, variation amplitude, temporal resolution and measurement accuracy. Furthermore, the cross-correlation analysis is always applied to the whole light curve instead of the trough part. If the light curve has other features, lags at troughs could be offset.

\section{Conclusion}\label{conclusion}
We monitored BL Lac object S5 0716+714 on seven nights from 2013 to 2016. Several telescopes were utilised for multi-colour quasi-simultaneous observations with high temporal resolution. The main results of our observation are as follow:
\begin{enumerate}
\item We use two statistical methods and IDVs are detected on all seven nights.
\item During our observations, the object turned to the brightest ($R=12.51$) on JD 2457437 and the faintest ($R=14.15$) on JD 2456322. The maximum intra-day variation is 0.15 mags in the $R$ band.
\item Achromatic and mild BWB intra-day spectral behaviours were observed. In the long term, the colour variation doesn't follow the BWB trend strictly.
\item On JD 2457315, a $\sim15$ minutes inter-band lag is detected by two independent methods.
\item We compare the lag with that of Wu12, and propose a hypothesis that this kind of inter-band lags at troughs between two flares could be produced due to the inconsistency of variation timescales at different wavelengths.
\end{enumerate}

\section*{Acknowledgements}
We thank the referee for important comments and constructive suggestions for improving our manuscript. This work has been supported by the National Natural Science Foundation of China grants U1531242.

\bibliography{0716_2018}

\begin{thebibliography}{}
\makeatletter
\relax
\def\mn@urlcharsother{\let\do\@makeother \do\$\do\&\do\#\do\^\do\_\do\%\do\~}
\def\mn@doi{\begingroup\mn@urlcharsother \@ifnextchar [ {\mn@doi@}
  {\mn@doi@[]}}
\def\mn@doi@[#1]#2{\def\@tempa{#1}\ifx\@tempa\@empty \href
  {http://dx.doi.org/#2} {doi:#2}\else \href {http://dx.doi.org/#2} {#1}\fi
  \endgroup}
\def\mn@eprint#1#2{\mn@eprint@#1:#2::\@nil}
\def\mn@eprint@arXiv#1{\href {http://arxiv.org/abs/#1} {{\tt arXiv:#1}}}
\def\mn@eprint@dblp#1{\href {http://dblp.uni-trier.de/rec/bibtex/#1.xml}
  {dblp:#1}}
\def\mn@eprint@#1:#2:#3:#4\@nil{\def\@tempa {#1}\def\@tempb {#2}\def\@tempc
  {#3}\ifx \@tempc \@empty \let \@tempc \@tempb \let \@tempb \@tempa \fi \ifx
  \@tempb \@empty \def\@tempb {arXiv}\fi \@ifundefined
  {mn@eprint@\@tempb}{\@tempb:\@tempc}{\expandafter \expandafter \csname
  mn@eprint@\@tempb\endcsname \expandafter{\@tempc}}}

\bibitem[\protect\citeauthoryear{{Abdo} et~al.,}{{Abdo}
  et~al.}{2010}]{2010ApJ...716...30A}
{Abdo} A.~A.,  et~al., 2010, \mn@doi [\apj] {10.1088/0004-637X/716/1/30}, \href
  {http://adsabs.harvard.edu/abs/2010ApJ...716...30A} {716, 30}

\bibitem[\protect\citeauthoryear{{Agarwal} et~al.,}{{Agarwal}
  et~al.}{2016}]{2016MNRAS.455..680A}
{Agarwal} A.,  et~al., 2016, \mn@doi [\mnras] {10.1093/mnras/stv2345}, \href
  {http://adsabs.harvard.edu/abs/2016MNRAS.455..680A} {455, 680}

\bibitem[\protect\citeauthoryear{{Alexander}}{{Alexander}}{1997}]{1997ASSL..218..163A}
{Alexander} T.,  1997, in {Maoz} D.,  {Sternberg} A.,   {Leibowitz} E.~M.,
  eds,  Astrophysics and Space Science Library Vol. 218, Astronomical Time
  Series. p.~163, \mn@doi{10.1007/978-94-015-8941-3_14}

\bibitem[\protect\citeauthoryear{{Alexander}}{{Alexander}}{2013}]{2013arXiv1302.1508A}
{Alexander} T.,  2013, preprint, \href
  {http://adsabs.harvard.edu/abs/2013arXiv1302.1508A} {} (\mn@eprint {arXiv}
  {1302.1508})

\bibitem[\protect\citeauthoryear{{Bachev}, {Strigachev}  \& {Semkov}}{{Bachev}
  et~al.}{2005}]{2005MNRAS.358..774B}
{Bachev} R.,  {Strigachev} A.,   {Semkov} E.,  2005, \mn@doi [\mnras]
  {10.1111/j.1365-2966.2005.08708.x}, \href
  {http://adsabs.harvard.edu/abs/2005MNRAS.358..774B} {358, 774}

\bibitem[\protect\citeauthoryear{{Bhatta} et~al.,}{{Bhatta}
  et~al.}{2013}]{2013A&A...558A..92B}
{Bhatta} G.,  et~al., 2013, \mn@doi [\aap] {10.1051/0004-6361/201220236}, \href
  {http://adsabs.harvard.edu/abs/2013A%26A...558A..92B} {558, A92}

\bibitem[\protect\citeauthoryear{{Bhatta} et~al.,}{{Bhatta}
  et~al.}{2016}]{2016ApJ...831...92B}
{Bhatta} G.,  et~al., 2016, \mn@doi [\apj] {10.3847/0004-637X/831/1/92}, \href
  {http://adsabs.harvard.edu/abs/2016ApJ...831...92B} {831, 92}

\bibitem[\protect\citeauthoryear{{Bonning} et~al.,}{{Bonning}
  et~al.}{2012}]{2012ApJ...756...13B}
{Bonning} E.,  et~al., 2012, \mn@doi [\apj] {10.1088/0004-637X/756/1/13}, \href
  {http://adsabs.harvard.edu/abs/2012ApJ...756...13B} {756, 13}

\bibitem[\protect\citeauthoryear{{B{\"o}ttcher} \& {Dermer}}{{B{\"o}ttcher} \&
  {Dermer}}{2010}]{2010ApJ...711..445B}
{B{\"o}ttcher} M.,  {Dermer} C.~D.,  2010, \mn@doi [\apj]
  {10.1088/0004-637X/711/1/445}, \href
  {http://adsabs.harvard.edu/abs/2010ApJ...711..445B} {711, 445}

\bibitem[\protect\citeauthoryear{{B{\"o}ttcher} et~al.,}{{B{\"o}ttcher}
  et~al.}{2003}]{2003ApJ...596..847B}
{B{\"o}ttcher} M.,  et~al., 2003, \mn@doi [\apj] {10.1086/378156}, \href
  {http://adsabs.harvard.edu/abs/2003ApJ...596..847B} {596, 847}

\bibitem[\protect\citeauthoryear{{Carini}, {Walters}  \& {Hopper}}{{Carini}
  et~al.}{2011}]{2011AJ....141...49C}
{Carini} M.~T.,  {Walters} R.,   {Hopper} L.,  2011, \mn@doi [\aj]
  {10.1088/0004-6256/141/2/49}, \href
  {http://adsabs.harvard.edu/abs/2011AJ....141...49C} {141, 49}

\bibitem[\protect\citeauthoryear{{Chiaberge} \& {Ghisellini}}{{Chiaberge} \&
  {Ghisellini}}{1999}]{1999MNRAS.306..551C}
{Chiaberge} M.,  {Ghisellini} G.,  1999, \mn@doi [\mnras]
  {10.1046/j.1365-8711.1999.02538.x}, \href
  {http://adsabs.harvard.edu/abs/1999MNRAS.306..551C} {306, 551}

\bibitem[\protect\citeauthoryear{{Dai}, {Wu}, {Zhu}, {Zhou}, {Ma}, {Yuan}  \&
  {Wang}}{{Dai} et~al.}{2013}]{2013ApJS..204...22D}
{Dai} Y.,  {Wu} J.,  {Zhu} Z.-H.,  {Zhou} X.,  {Ma} J.,  {Yuan} Q.,   {Wang}
  L.,  2013, \mn@doi [\apjs] {10.1088/0067-0049/204/2/22}, \href
  {http://adsabs.harvard.edu/abs/2013ApJS..204...22D} {204, 22}

\bibitem[\protect\citeauthoryear{{Danforth}, {Nalewajko}, {France}  \&
  {Keeney}}{{Danforth} et~al.}{2013}]{2013ApJ...764...57D}
{Danforth} C.~W.,  {Nalewajko} K.,  {France} K.,   {Keeney} B.~A.,  2013,
  \mn@doi [\apj] {10.1088/0004-637X/764/1/57}, \href
  {http://adsabs.harvard.edu/abs/2013ApJ...764...57D} {764, 57}

\bibitem[\protect\citeauthoryear{{Dermer}}{{Dermer}}{1998}]{1998ApJ...501L.157D}
{Dermer} C.~D.,  1998, \mn@doi [\apjl] {10.1086/311467}, \href
  {http://adsabs.harvard.edu/abs/1998ApJ...501L.157D} {501, L157}

\bibitem[\protect\citeauthoryear{{Edelson} \& {Krolik}}{{Edelson} \&
  {Krolik}}{1988}]{1988ApJ...333..646E}
{Edelson} R.~A.,  {Krolik} J.~H.,  1988, \mn@doi [\apj] {10.1086/166773}, \href
  {http://adsabs.harvard.edu/abs/1988ApJ...333..646E} {333, 646}

\bibitem[\protect\citeauthoryear{{Fiorucci}, {Ciprini}  \& {Tosti}}{{Fiorucci}
  et~al.}{2004}]{2004A&A...419...25F}
{Fiorucci} M.,  {Ciprini} S.,   {Tosti} G.,  2004, \mn@doi [\aap]
  {10.1051/0004-6361:20034218}, \href
  {http://adsabs.harvard.edu/abs/2004A%26A...419...25F} {419, 25}

\bibitem[\protect\citeauthoryear{{Fossati} et~al.,}{{Fossati}
  et~al.}{2000a}]{2000ApJ...541..153F}
{Fossati} G.,  et~al., 2000a, \mn@doi [\apj] {10.1086/309422}, \href
  {http://adsabs.harvard.edu/abs/2000ApJ...541..153F} {541, 153}

\bibitem[\protect\citeauthoryear{{Fossati} et~al.,}{{Fossati}
  et~al.}{2000b}]{2000ApJ...541..166F}
{Fossati} G.,  et~al., 2000b, \mn@doi [\apj] {10.1086/309430}, \href
  {http://adsabs.harvard.edu/abs/2000ApJ...541..166F} {541, 166}

\bibitem[\protect\citeauthoryear{{Garcia}, {Sodr{\'e}}, {Jablonski}  \&
  {Terlevich}}{{Garcia} et~al.}{1999}]{1999MNRAS.309..803G}
{Garcia} A.,  {Sodr{\'e}} L.,  {Jablonski} F.~J.,   {Terlevich} R.~J.,  1999,
  \mn@doi [\mnras] {10.1046/j.1365-8711.1999.02884.x}, \href
  {http://adsabs.harvard.edu/abs/1999MNRAS.309..803G} {309, 803}

\bibitem[\protect\citeauthoryear{{Gaur}, {Gupta}  \& {Wiita}}{{Gaur}
  et~al.}{2012a}]{2012AJ....143...23G}
{Gaur} H.,  {Gupta} A.~C.,   {Wiita} P.~J.,  2012a, \mn@doi [\aj]
  {10.1088/0004-6256/143/1/23}, \href
  {http://adsabs.harvard.edu/abs/2012AJ....143...23G} {143, 23}

\bibitem[\protect\citeauthoryear{{Gaur} et~al.,}{{Gaur}
  et~al.}{2012b}]{2012MNRAS.420.3147G}
{Gaur} H.,  et~al., 2012b, \mn@doi [\mnras] {10.1111/j.1365-2966.2011.20243.x},
  \href {http://adsabs.harvard.edu/abs/2012MNRAS.420.3147G} {420, 3147}

\bibitem[\protect\citeauthoryear{{Gaur} et~al.,}{{Gaur}
  et~al.}{2012c}]{2012MNRAS.425.3002G}
{Gaur} H.,  et~al., 2012c, \mn@doi [\mnras] {10.1111/j.1365-2966.2012.21583.x},
  \href {http://adsabs.harvard.edu/abs/2012MNRAS.425.3002G} {425, 3002}

\bibitem[\protect\citeauthoryear{{Ghisellini} et~al.,}{{Ghisellini}
  et~al.}{1997}]{1997A&A...327...61G}
{Ghisellini} G.,  et~al., 1997, \aap, \href
  {http://adsabs.harvard.edu/abs/1997A%26A...327...61G} {327, 61}

\bibitem[\protect\citeauthoryear{{Gopal-Krishna}, {Stalin}, {Sagar}  \&
  {Wiita}}{{Gopal-Krishna} et~al.}{2003}]{2003ApJ...586L..25G}
{Gopal-Krishna} {Stalin} C.~S.,  {Sagar} R.,   {Wiita} P.~J.,  2003, \mn@doi
  [\apjl] {10.1086/374655}, \href
  {http://adsabs.harvard.edu/abs/2003ApJ...586L..25G} {586, L25}

\bibitem[\protect\citeauthoryear{{Goyal}, {Mhaskey}, {Gopal-Krishna}, {Wiita},
  {Stalin}  \& {Sagar}}{{Goyal} et~al.}{2013}]{2013JApA...34..273G}
{Goyal} A.,  {Mhaskey} M.,  {Gopal-Krishna} {Wiita} P.~J.,  {Stalin} C.~S.,
  {Sagar} R.,  2013, \mn@doi [Journal of Astrophysics and Astronomy]
  {10.1007/s12036-013-9183-7}, \href
  {http://adsabs.harvard.edu/abs/2013JApA...34..273G} {34, 273}

\bibitem[\protect\citeauthoryear{{Gu}, {Lee}, {Pak}, {Yim}  \& {Fletcher}}{{Gu}
  et~al.}{2006}]{2006A&A...450...39G}
{Gu} M.~F.,  {Lee} C.-U.,  {Pak} S.,  {Yim} H.~S.,   {Fletcher} A.~B.,  2006,
  \mn@doi [\aap] {10.1051/0004-6361:20054271}, \href
  {http://adsabs.harvard.edu/abs/2006A%26A...450...39G} {450, 39}

\bibitem[\protect\citeauthoryear{{Gupta}}{{Gupta}}{2018}]{2018Galax...6....1G}
{Gupta} A.,  2018, \mn@doi [Galaxies] {10.3390/galaxies6010001}, \href
  {http://adsabs.harvard.edu/abs/2018Galax...6....1G} {6, 1}

\bibitem[\protect\citeauthoryear{{Gupta} \& {Joshi}}{{Gupta} \&
  {Joshi}}{2005}]{2005A&A...440..855G}
{Gupta} A.~C.,  {Joshi} U.~C.,  2005, \mn@doi [\aap]
  {10.1051/0004-6361:20042370}, \href
  {http://adsabs.harvard.edu/abs/2005A%26A...440..855G} {440, 855}

\bibitem[\protect\citeauthoryear{{Gupta}, {Fan}, {Bai}  \& {Wagner}}{{Gupta}
  et~al.}{2008}]{2008AJ....135.1384G}
{Gupta} A.~C.,  {Fan} J.~H.,  {Bai} J.~M.,   {Wagner} S.~J.,  2008, \mn@doi
  [\aj] {10.1088/0004-6256/135/4/1384}, \href
  {http://adsabs.harvard.edu/abs/2008AJ....135.1384G} {135, 1384}

\bibitem[\protect\citeauthoryear{{Heidt} \& {Wagner}}{{Heidt} \&
  {Wagner}}{1996}]{1996A&A...305...42H}
{Heidt} J.,  {Wagner} S.~J.,  1996, \aap, \href
  {http://adsabs.harvard.edu/abs/1996A%26A...305...42H} {305, 42}

\bibitem[\protect\citeauthoryear{{Hong}, {Xiong}  \& {Bai}}{{Hong}
  et~al.}{2017}]{2017AJ....154...42H}
{Hong} S.,  {Xiong} D.,   {Bai} J.,  2017, \mn@doi [\aj]
  {10.3847/1538-3881/aa799a}, \href
  {http://adsabs.harvard.edu/abs/2017AJ....154...42H} {154, 42}

\bibitem[\protect\citeauthoryear{{Hong}, {Xiong}  \& {Bai}}{{Hong}
  et~al.}{2018}]{2018AJ....155...31H}
{Hong} S.,  {Xiong} D.,   {Bai} J.,  2018, \mn@doi [\aj]
  {10.3847/1538-3881/aa9d89}, \href
  {http://adsabs.harvard.edu/abs/2018AJ....155...31H} {155, 31}

\bibitem[\protect\citeauthoryear{{Howell}, {Mitchell}  \& {Warnock}}{{Howell}
  et~al.}{1988}]{1988AJ.....95..247H}
{Howell} S.~B.,  {Mitchell} K.~J.,   {Warnock} III A.,  1988, \mn@doi [\aj]
  {10.1086/114634}, \href {http://adsabs.harvard.edu/abs/1988AJ.....95..247H}
  {95, 247}

\bibitem[\protect\citeauthoryear{{Hu}, {Chen}, {Guo}, {Jiang}  \& {Li}}{{Hu}
  et~al.}{2014}]{2014MNRAS.443.2940H}
{Hu} S.~M.,  {Chen} X.,  {Guo} D.~F.,  {Jiang} Y.~G.,   {Li} K.,  2014, \mn@doi
  [\mnras] {10.1093/mnras/stu1373}, \href
  {http://adsabs.harvard.edu/abs/2014MNRAS.443.2940H} {443, 2940}

\bibitem[\protect\citeauthoryear{{Ikejiri} et~al.,}{{Ikejiri}
  et~al.}{2011}]{2011PASJ...63..639I}
{Ikejiri} Y.,  et~al., 2011, \mn@doi [\pasj] {10.1093/pasj/63.3.327}, \href
  {http://adsabs.harvard.edu/abs/2011PASJ...63..639I} {63, 639}

\bibitem[\protect\citeauthoryear{{Kataoka}, {Takahashi}, {Makino}, {Inoue},
  {Madejski}, {Tashiro}, {Urry}  \& {Kubo}}{{Kataoka}
  et~al.}{2000}]{2000ApJ...528..243K}
{Kataoka} J.,  {Takahashi} T.,  {Makino} F.,  {Inoue} S.,  {Madejski} G.~M.,
  {Tashiro} M.,  {Urry} C.~M.,   {Kubo} H.,  2000, \mn@doi [\apj]
  {10.1086/308154}, \href {http://adsabs.harvard.edu/abs/2000ApJ...528..243K}
  {528, 243}

\bibitem[\protect\citeauthoryear{{Kirk}, {Rieger}  \& {Mastichiadis}}{{Kirk}
  et~al.}{1998}]{1998A&A...333..452K}
{Kirk} J.~G.,  {Rieger} F.~M.,   {Mastichiadis} A.,  1998, \aap, \href
  {http://adsabs.harvard.edu/abs/1998A%26A...333..452K} {333, 452}

\bibitem[\protect\citeauthoryear{{Liao}, {Bai}, {Liu}, {Weng}, {Chen}  \&
  {Li}}{{Liao} et~al.}{2014}]{2014ApJ...783...83L}
{Liao} N.~H.,  {Bai} J.~M.,  {Liu} H.~T.,  {Weng} S.~S.,  {Chen} L.,   {Li} F.,
   2014, \mn@doi [\apj] {10.1088/0004-637X/783/2/83}, \href
  {http://adsabs.harvard.edu/abs/2014ApJ...783...83L} {783, 83}

\bibitem[\protect\citeauthoryear{{Liu} et~al.,}{{Liu}
  et~al.}{2017}]{2017MNRAS.469.2457L}
{Liu} X.,  et~al., 2017, \mn@doi [\mnras] {10.1093/mnras/stx1062}, \href
  {http://adsabs.harvard.edu/abs/2017MNRAS.469.2457L} {469, 2457}

\bibitem[\protect\citeauthoryear{{Man}, {Zhang}, {Wu}  \& {Yuan}}{{Man}
  et~al.}{2016}]{2016MNRAS.456.3168M}
{Man} Z.,  {Zhang} X.,  {Wu} J.,   {Yuan} Q.,  2016, \mn@doi [\mnras]
  {10.1093/mnras/stv2879}, \href
  {http://adsabs.harvard.edu/abs/2016MNRAS.456.3168M} {456, 3168}

\bibitem[\protect\citeauthoryear{{Miller}, {Carini}  \& {Goodrich}}{{Miller}
  et~al.}{1989}]{1989Natur.337..627M}
{Miller} H.~R.,  {Carini} M.~T.,   {Goodrich} B.~D.,  1989, \mn@doi [\nat]
  {10.1038/337627a0}, \href {http://adsabs.harvard.edu/abs/1989Natur.337..627M}
  {337, 627}

\bibitem[\protect\citeauthoryear{{Montagni}, {Maselli}, {Massaro}, {Nesci},
  {Sclavi}  \& {Maesano}}{{Montagni} et~al.}{2006}]{2006A&A...451..435M}
{Montagni} F.,  {Maselli} A.,  {Massaro} E.,  {Nesci} R.,  {Sclavi} S.,
  {Maesano} M.,  2006, \mn@doi [\aap] {10.1051/0004-6361:20053874}, \href
  {http://adsabs.harvard.edu/abs/2006A%26A...451..435M} {451, 435}

\bibitem[\protect\citeauthoryear{{Nesci}, {Massaro}, {Rossi}, {Sclavi},
  {Maesano}  \& {Montagni}}{{Nesci} et~al.}{2005}]{2005AJ....130.1466N}
{Nesci} R.,  {Massaro} E.,  {Rossi} C.,  {Sclavi} S.,  {Maesano} M.,
  {Montagni} F.,  2005, \mn@doi [\aj] {10.1086/444538}, \href
  {http://adsabs.harvard.edu/abs/2005AJ....130.1466N} {130, 1466}

\bibitem[\protect\citeauthoryear{{Nilsson}, {Pursimo}, {Sillanp{\"a}{\"a}},
  {Takalo}  \& {Lindfors}}{{Nilsson} et~al.}{2008}]{2008A&A...487L..29N}
{Nilsson} K.,  {Pursimo} T.,  {Sillanp{\"a}{\"a}} A.,  {Takalo} L.~O.,
  {Lindfors} E.,  2008, \mn@doi [\aap] {10.1051/0004-6361:200810310}, \href
  {http://adsabs.harvard.edu/abs/2008A%26A...487L..29N} {487, L29}

\bibitem[\protect\citeauthoryear{{Peterson}, {Wanders}, {Horne}, {Collier},
  {Alexander}, {Kaspi}  \& {Maoz}}{{Peterson}
  et~al.}{1998}]{1998PASP..110..660P}
{Peterson} B.~M.,  {Wanders} I.,  {Horne} K.,  {Collier} S.,  {Alexander} T.,
  {Kaspi} S.,   {Maoz} D.,  1998, \mn@doi [\pasp] {10.1086/316177}, \href
  {http://adsabs.harvard.edu/abs/1998PASP..110..660P} {110, 660}

\bibitem[\protect\citeauthoryear{{Peterson} et~al.,}{{Peterson}
  et~al.}{2004}]{2004ApJ...613..682P}
{Peterson} B.~M.,  et~al., 2004, \mn@doi [\apj] {10.1086/423269}, \href
  {http://adsabs.harvard.edu/abs/2004ApJ...613..682P} {613, 682}

\bibitem[\protect\citeauthoryear{{Poon}, {Fan}  \& {Fu}}{{Poon}
  et~al.}{2009}]{2009ApJS..185..511P}
{Poon} H.,  {Fan} J.~H.,   {Fu} J.~N.,  2009, \mn@doi [\apjs]
  {10.1088/0067-0049/185/2/511}, \href
  {http://adsabs.harvard.edu/abs/2009ApJS..185..511P} {185, 511}

\bibitem[\protect\citeauthoryear{{Qian}, {Tao}  \& {Fan}}{{Qian}
  et~al.}{2000}]{2000PASJ...52.1075Q}
{Qian} B.,  {Tao} J.,   {Fan} J.,  2000, \mn@doi [\pasj]
  {10.1093/pasj/52.6.1075}, \href
  {http://adsabs.harvard.edu/abs/2000PASJ...52.1075Q} {52, 1075}

\bibitem[\protect\citeauthoryear{{Raiteri} et~al.,}{{Raiteri}
  et~al.}{2003}]{2003A&A...402..151R}
{Raiteri} C.~M.,  et~al., 2003, \mn@doi [\aap] {10.1051/0004-6361:20030256},
  \href {http://adsabs.harvard.edu/abs/2003A%26A...402..151R} {402, 151}

\bibitem[\protect\citeauthoryear{{Raiteri} et~al.,}{{Raiteri}
  et~al.}{2008}]{2008A&A...491..755R}
{Raiteri} C.~M.,  et~al., 2008, \mn@doi [\aap] {10.1051/0004-6361:200810869},
  \href {http://adsabs.harvard.edu/abs/2008A%26A...491..755R} {491, 755}

\bibitem[\protect\citeauthoryear{{Rani} et~al.,}{{Rani}
  et~al.}{2010a}]{2010MNRAS.404.1992R}
{Rani} B.,  et~al., 2010a, \mn@doi [\mnras] {10.1111/j.1365-2966.2010.16419.x},
  \href {http://adsabs.harvard.edu/abs/2010MNRAS.404.1992R} {404, 1992}

\bibitem[\protect\citeauthoryear{{Rani}, {Gupta}, {Joshi}, {Ganesh}  \&
  {Wiita}}{{Rani} et~al.}{2010b}]{2010ApJ...719L.153R}
{Rani} B.,  {Gupta} A.~C.,  {Joshi} U.~C.,  {Ganesh} S.,   {Wiita} P.~J.,
  2010b, \mn@doi [\apjl] {10.1088/2041-8205/719/2/L153}, \href
  {http://adsabs.harvard.edu/abs/2010ApJ...719L.153R} {719, L153}

\bibitem[\protect\citeauthoryear{{Rani} et~al.,}{{Rani}
  et~al.}{2013}]{2013A&A...552A..11R}
{Rani} B.,  et~al., 2013, \mn@doi [\aap] {10.1051/0004-6361/201321058}, \href
  {http://adsabs.harvard.edu/abs/2013A%26A...552A..11R} {552, A11}

\bibitem[\protect\citeauthoryear{{Spada}, {Ghisellini}, {Lazzati}  \&
  {Celotti}}{{Spada} et~al.}{2001}]{2001MNRAS.325.1559S}
{Spada} M.,  {Ghisellini} G.,  {Lazzati} D.,   {Celotti} A.,  2001, \mn@doi
  [\mnras] {10.1046/j.1365-8711.2001.04557.x}, \href
  {http://adsabs.harvard.edu/abs/2001MNRAS.325.1559S} {325, 1559}

\bibitem[\protect\citeauthoryear{{Stalin}, {Gopal-Krishna}, {Sagar}  \&
  {Wiita}}{{Stalin} et~al.}{2004}]{2004MNRAS.350..175S}
{Stalin} C.~S.,  {Gopal-Krishna} {Sagar} R.,   {Wiita} P.~J.,  2004, \mn@doi
  [\mnras] {10.1111/j.1365-2966.2004.07631.x}, \href
  {http://adsabs.harvard.edu/abs/2004MNRAS.350..175S} {350, 175}

\bibitem[\protect\citeauthoryear{{Stalin}, {Gopal-Krishna}, {Sagar}, {Wiita},
  {Mohan}  \& {Pandey}}{{Stalin} et~al.}{2006}]{2006MNRAS.366.1337S}
{Stalin} C.~S.,  {Gopal-Krishna} {Sagar} R.,  {Wiita} P.~J.,  {Mohan} V.,
  {Pandey} A.~K.,  2006, \mn@doi [\mnras] {10.1111/j.1365-2966.2005.09939.x},
  \href {http://adsabs.harvard.edu/abs/2006MNRAS.366.1337S} {366, 1337}

\bibitem[\protect\citeauthoryear{{Takahashi} et~al.,}{{Takahashi}
  et~al.}{1996}]{1996ApJ...470L..89T}
{Takahashi} T.,  et~al., 1996, \mn@doi [\apjl] {10.1086/310302}, \href
  {http://adsabs.harvard.edu/abs/1996ApJ...470L..89T} {470, L89}

\bibitem[\protect\citeauthoryear{{Urry} \& {Padovani}}{{Urry} \&
  {Padovani}}{1995}]{1995PASP..107..803U}
{Urry} C.~M.,  {Padovani} P.,  1995, \mn@doi [\pasp] {10.1086/133630}, \href
  {http://adsabs.harvard.edu/abs/1995PASP..107..803U} {107, 803}

\bibitem[\protect\citeauthoryear{{Villata}, {Raiteri}, {Lanteri}, {Sobrito}  \&
  {Cavallone}}{{Villata} et~al.}{1998}]{1998A&AS..130..305V}
{Villata} M.,  {Raiteri} C.~M.,  {Lanteri} L.,  {Sobrito} G.,   {Cavallone} M.,
   1998, \mn@doi [\aaps] {10.1051/aas:1998415}, \href
  {http://adsabs.harvard.edu/abs/1998A%26AS..130..305V} {130, 305}

\bibitem[\protect\citeauthoryear{{Villata} et~al.,}{{Villata}
  et~al.}{2000}]{2000A&A...363..108V}
{Villata} M.,  et~al., 2000, \aap, \href
  {http://adsabs.harvard.edu/abs/2000A%26A...363..108V} {363, 108}

\bibitem[\protect\citeauthoryear{{Villata} et~al.,}{{Villata}
  et~al.}{2008}]{2008A&A...481L..79V}
{Villata} M.,  et~al., 2008, \mn@doi [\aap] {10.1051/0004-6361:200809552},
  \href {http://adsabs.harvard.edu/abs/2008A%26A...481L..79V} {481, L79}

\bibitem[\protect\citeauthoryear{{Villata} et~al.,}{{Villata}
  et~al.}{2009}]{2009A&A...501..455V}
{Villata} M.,  et~al., 2009, \mn@doi [\aap] {10.1051/0004-6361/200912065},
  \href {http://adsabs.harvard.edu/abs/2009A%26A...501..455V} {501, 455}

\bibitem[\protect\citeauthoryear{{Wagner} et~al.,}{{Wagner}
  et~al.}{1996}]{1996AJ....111.2187W}
{Wagner} S.~J.,  et~al., 1996, \mn@doi [\aj] {10.1086/117954}, \href
  {http://adsabs.harvard.edu/abs/1996AJ....111.2187W} {111, 2187}

\bibitem[\protect\citeauthoryear{{Wierzcholska}, {Ostrowski}, {Stawarz},
  {Wagner}  \& {Hauser}}{{Wierzcholska} et~al.}{2015}]{2015A&A...573A..69W}
{Wierzcholska} A.,  {Ostrowski} M.,  {Stawarz} {\L}.,  {Wagner} S.,   {Hauser}
  M.,  2015, \mn@doi [\aap] {10.1051/0004-6361/201423967}, \href
  {http://adsabs.harvard.edu/abs/2015A%26A...573A..69W} {573, A69}

\bibitem[\protect\citeauthoryear{{Wu}, {Peng}, {Zhou}, {Ma}, {Jiang}  \&
  {Chen}}{{Wu} et~al.}{2005}]{2005AJ....129.1818W}
{Wu} J.,  {Peng} B.,  {Zhou} X.,  {Ma} J.,  {Jiang} Z.,   {Chen} J.,  2005,
  \mn@doi [\aj] {10.1086/428599}, \href
  {http://adsabs.harvard.edu/abs/2005AJ....129.1818W} {129, 1818}

\bibitem[\protect\citeauthoryear{{Wu}, {Zhou}, {Ma}, {Wu}, {Jiang}  \&
  {Chen}}{{Wu} et~al.}{2007}]{2007AJ....133.1599W}
{Wu} J.,  {Zhou} X.,  {Ma} J.,  {Wu} Z.,  {Jiang} Z.,   {Chen} J.,  2007,
  \mn@doi [\aj] {10.1086/511773}, \href
  {http://adsabs.harvard.edu/abs/2007AJ....133.1599W} {133, 1599}

\bibitem[\protect\citeauthoryear{{Wu}, {B{\"o}ttcher}, {Zhou}, {He}, {Ma}  \&
  {Jiang}}{{Wu} et~al.}{2012}]{2012AJ....143..108W}
{Wu} J.,  {B{\"o}ttcher} M.,  {Zhou} X.,  {He} X.,  {Ma} J.,   {Jiang} Z.,
  2012, \mn@doi [\aj] {10.1088/0004-6256/143/5/108}, \href
  {http://adsabs.harvard.edu/abs/2012AJ....143..108W} {143, 108}

\bibitem[\protect\citeauthoryear{{Zhang} et~al.,}{{Zhang}
  et~al.}{1999}]{1999ApJ...527..719Z}
{Zhang} Y.~H.,  et~al., 1999, \mn@doi [\apj] {10.1086/308116}, \href
  {http://adsabs.harvard.edu/abs/1999ApJ...527..719Z} {527, 719}

\bibitem[\protect\citeauthoryear{{de Diego}}{{de
  Diego}}{2010}]{2010AJ....139.1269D}
{de Diego} J.~A.,  2010, \mn@doi [\aj] {10.1088/0004-6256/139/3/1269}, \href
  {http://adsabs.harvard.edu/abs/2010AJ....139.1269D} {139, 1269}

\bibitem[\protect\citeauthoryear{{de Diego}, {Polednikova}, {Bongiovanni},
  {P{\'e}rez Garc{\'{\i}}a}, {De Leo}, {Verdugo}  \& {Cepa}}{{de Diego}
  et~al.}{2015}]{2015AJ....150...44D}
{de Diego} J.~A.,  {Polednikova} J.,  {Bongiovanni} A.,  {P{\'e}rez
  Garc{\'{\i}}a} A.~M.,  {De Leo} M.~A.,  {Verdugo} T.,   {Cepa} J.,  2015,
  \mn@doi [\aj] {10.1088/0004-6256/150/2/44}, \href
  {http://adsabs.harvard.edu/abs/2015AJ....150...44D} {150, 44}

\makeatother
\end{thebibliography}
\bibliographystyle{mnras}

\bsp
\label{lastpage}

\end{document}